\documentclass[aps,prb,reprint,preprintnumbers,superscriptaddress,amsmath,amssymb,bibnotes,longbibliography]{revtex4-2}

\usepackage{graphicx}
\usepackage{amsmath}
\usepackage{dcolumn}
\usepackage{epstopdf}
\usepackage{bm,multirow,threeparttable,soul}
\usepackage{flushend}
\usepackage[pdfstartview=FitH, CJKbookmarks=true, bookmarksnumbered=true, bookmarksopen=true, colorlinks=true, pdfborder={0 0 1}, linkcolor=blue, anchorcolor=blue, citecolor=blue,urlcolor=blue,breaklinks=true]{hyperref}

\usepackage[mathlines]{lineno}

\begin{document}

\preprint{APS/123-QED}

\title{\textbf{Nodeless superconductivity in 4$\texorpdfstring{H_{b}}{}$-TaS$\texorpdfstring{_2}{}$ with broken time reversal symmetry} 
}%

\author{Yuwei Zhou}
 \affiliation{Center for Correlated Matter and School of Physics, Zhejiang University, Hangzhou 310058, China}
 \author{Fanyu Meng}
 \affiliation{School of Physics and Beijing Key Laboratory of Opto-electronic Functional Materials \& Micro-nano Devices, Renmin University of China, Beijing 100872, China}
\affiliation{Key Laboratory of Quantum State Construction and Manipulation (Ministry of Educatoin), Renmin University of China, Beijing 100872, China}
\author{Yanen Huang}
 \affiliation{Center for Correlated Matter and School of Physics, Zhejiang University, Hangzhou 310058, China}
 \author{Jiawen Zhang}
 \affiliation{Center for Correlated Matter and School of Physics, Zhejiang University, Hangzhou 310058, China}
 \author{Jin Zhan}
 \affiliation{Center for Correlated Matter and School of Physics, Zhejiang University, Hangzhou 310058, China}
 \author{Ye Chen}
 \affiliation{Center for Correlated Matter and School of Physics, Zhejiang University, Hangzhou 310058, China}
\author{Yu Liu}
 \affiliation{Center for Correlated Matter and School of Physics, Zhejiang University, Hangzhou 310058, China}
 \author{Hechang Lei}
 \affiliation{School of Physics and Beijing Key Laboratory of Opto-electronic Functional Materials \& Micro-nano Devices, Renmin University of China, Beijing 100872, China}
\affiliation{Key Laboratory of Quantum State Construction and Manipulation (Ministry of Educatoin), Renmin University of China, Beijing 100872, China}
\author{Michael Smidman}
\email[Corresponding author: ]{msmidman@zju.edu.cn}
\affiliation{Center for Correlated Matter and School of Physics, Zhejiang University, Hangzhou 310058, China}
\author{Huiqiu Yuan}
\email[Corresponding author: ]{hqyuan@zju.edu.cn}
\affiliation{Center for Correlated Matter and School of Physics, Zhejiang University, Hangzhou 310058, China}
\affiliation{Institute for Advanced Study in Physics, Zhejiang University, Hangzhou 310058, China}
\affiliation{Institute of Fundamental and Transdisciplinary Research, Zhejiang University, Hangzhou 310058, China}
\affiliation{State Key Laboratory of Silicon and Advanced Semiconductor Materials, Zhejiang University, Hangzhou 310058, China}

\date{\today}

\begin{abstract}
The transition metal dichalcogenide 4$\texorpdfstring{H_{b}}{}$-TaS$\texorpdfstring{_2}{}$ exhibits characteristics of topological edge modes and two-component superconductivity with time-reversal symmetry breaking (TRSB). The nature of the superconducting order parameter is a crucial issue that requires experimental investigation. Here, we report measurements of the magnetic penetration depth using a tunnel-diode-oscillator based technique, as well as the specific heat. Both the specific heat and the change in magnetic penetration depth ($\texorpdfstring{\Delta \lambda(T)}{}$) display an exponentially-activated temperature dependence, providing evidence for nodeless superconductivity in 4$\texorpdfstring{H_{b}}{}$-TaS$\texorpdfstring{_2}{}$. Moreover, the deduced superfluid density can be well described by a two-gap $\texorpdfstring{s}{}$-wave model, and such multigap superconductivity is consistent with there being multiple bands crossing the Fermi energy. These results constrain the possible pairing symmetries of 4$\texorpdfstring{H_{b}}{}$-TaS$\texorpdfstring{_2}{}$.

\begin{description}
\item[PACS number(s)]

\end{description}
\end{abstract}

\maketitle
\section{INTRODUCTION}
Time-reversal symmetry breaking (TRSB) in superconductors is manifested by the appearance of spontaneous magnetic fields in the superconducting state. Such an additional symmetry breaking at $T_\mathrm{c}$ generally requires a degenerate instability channel, leading to a multi-component order parameter, often with nodal superconducting gaps. These order parameters correspond to an unconventional superconducting state~\cite{sigrist1991phenomenological,ghosh2020recent}, as observed in some strongly correlated electronic systems~\cite{luke1993muon,luke1998time,xia2006high,heffner1990new,schemm2015evidence,schemm2014observation}. More recently, evidence for superconductivity with TRSB was observed in some weakly correlated materials, such as LaNiC$_2$~\cite{hillier2009evidence}, LaNiGa$_2$~\cite{hillier2012nonunitary,weng2016two}, Lu$_{5-x}$Rh$_6$Sn$_{18+x}$~\cite{bhattacharyya2015broken,wang2021nodeless}, CaPtAs~\cite{xie2020captas,shang2020simultaneous} and several Re-based superconductors~\cite{singh2017time,pang2018fully,singh2018time}. Thus, whether superconductivity with TRSB necessarily originates from an unconventional pairing mechanism or can be realized through conventional electron-phonon pairing requires further investigation in more candidate materials.

The quasi-two-dimensional transition metal dichalcogenide materials exhibit intricate interplay between charge, lattice, and orbital degrees of freedom, giving rise to unusual properties, such as charge density wave phases, superconductivity and topological phases~\cite{wilson1975charge,wilson1974charge,manzeli20172d,liu2014coexistence}. Depending on the stacking arrangement and the unit cell size, diverse structural phases, including 1$T$, 2$H$, 3$R$, 4$H_{b}$, and 6$R$, are obtained~\cite{di1973preparation}. The natural van der Waals heterostructure 4$\texorpdfstring{H_{b}}{}$-TaS$\texorpdfstring{_2}{}$, consisting of alternating layers of octahedral 1$H$-TaS$_2$ and trigonal prismatic 1$T$-TaS$_2$, undergoes a superconducting transition at 2.9~K. 1$H$-TaS$_2$ exhibits superconductivity at 0.8~K in its bulk form~\cite{wilson1975charge}, whereas Mott physics plays a crucial role in both bulk and single-layer 1$T$-TaS$_2$~\cite{fei2022understanding,law20171t,lin2020scanning,sipos2008mott}. A chiral superconducting state has been proposed for 4$\texorpdfstring{H_{b}}{}$-TaS$\texorpdfstring{_2}{}$ based on muon spin relaxation ($\mu\mathrm{SR}$) measurements showing TRSB below the superconducting transition temperature~\cite{ribak2020chiral}. Moreover, 4$\texorpdfstring{H_{b}}{}$-TaS$\texorpdfstring{_2}{}$ has been proposed as a candidate for topological superconductivity based on scanning tunneling spectroscopy measurements, which exhibit signatures of edge modes and zero-bias states in vortex cores~\cite{nayak2021evidence}. Little-Parks effect measurements reveal $\pi$-shifts in the superconducting transition-temperature oscillations, providing possible evidence for a multi-component order parameter~\cite{almoalem2024observation}. 

The superconducting gap structure of 4$\texorpdfstring{H_{b}}{}$-TaS$\texorpdfstring{_2}{}$ still remains unclear. Theoretically, some symmetry-imposed nodal structures, such as those of the  B$_{1u}$ and E$_{2g}$ order parameters, have been suggested~\cite{fischer2015symmetry}. Experimentally, scanning tunneling microscopy (STM) measurements reveal a U-shaped superconducting gap but with a finite in-gap density of states~\cite{nayak2021evidence}. Moreover, a small residual linear term in zero field and an S-shaped field dependence are observed in thermal conductivity measurements~\cite{wang2024evidence}, and it was suggested that some Fermi surfaces are fully gapped while others remain gapless. Recent measurements of transverse-field (TF) $\mu\mathrm{SR}$ evidence fully gapped superconductivity, while the specific heat coefficient exhibits a finite intersect in the $T\rightarrow0$ limit, which is suggested to be an intrinsic feature in 4$\texorpdfstring{H_{b}}{}$-TaS$\texorpdfstring{_2}{}$~\cite{ribak2020chiral}. To better understand the superconducting order parameter of 4$\texorpdfstring{H_{b}}{}$-TaS$\texorpdfstring{_2}{}$, experiments sensitive to low-energy electronic excitations within the superconducting state are required.

In this article, we probe the gap symmetry of 4$H_{b}$-TaS$_{1.99}$Se$_{0.01}$ single crystals via measurements of the magnetic penetration depth change $\Delta \lambda(T)$ using a tunnel-diode oscillator (TDO) based technique, as well as the specific heat. The $\Delta \lambda(T)$ flattens at low temperatures and shows an exponentially activated behavior, which rules out the presence of nodes, demonstrating fully-gapped superconductivity. Furthermore, the deduced superfluid density and electronic specific heat data can be well fitted by a two-gap $s$-wave model. These results provide evidence for nodeless multigap superconductivity in 4$\texorpdfstring{H_{b}}{}$-TaS$\texorpdfstring{_2}{}$. 
  
\section{EXPERIMENTAL DETAILS}
Single crystals of 4$H_{b}$-TaS$_{1.99}$Se$_{0.01}$ were synthesized using the chemical vapor transport method, with 1\% Se doping to enhance the stability of the crystal structure, reduce the superconducting transition width, and significantly increase the superconducting volume fraction, as reported in previous studies~\cite{meng2024extreme,ribak2020chiral,persky2022magnetic,nayak2021evidence,almoalem2024observation,almoalem2024charge}. The electrical resistivity $\rho(T)$ and heat capacity $C(T)$ were measured using a Physical Property Measurement System (PPMS) down to 2~K and 0.4~K, respectively. Resistivity measurements were performed on sample (\#2) which was also measured using the TDO technique. Four platinum wires were attached to the sample surface using silver paint. The sample mass for the specific heat measurement was 11.8~mg. The magnetization measurements down to 1.8~K were performed using a Magnetic Property Measurement System (MPMS). The temperature dependence of the magnetic penetration depth shift, $\Delta \lambda(T) = \lambda(T) - \lambda(0)$, was measured using a TDO-based technique in a $^3$He cryostat down to 0.33~K, operating at approximately 7~MHz with a noise level as low as 0.15~Hz~\cite{van1975tunnel,prozorov2006magnetic}. $\Delta \lambda(T)$ can be obtained from the TDO frequency shift $\Delta f(T)$ through $\Delta \lambda(T) = G \Delta f(T)$, where $G$ is the calibration factor determined from the geometry of the coil and the sample~\cite{prozorov2000meissner,prozorov2018effective}. Three samples were oriented using an x-ray Laue, cut into regular shapes, and then separately mounted onto a sapphire rod, which was positioned without direct contact with the coil. The dimensions of the three samples are 725$\times$580$\times$50 $\mu$m$^3$ (\#1), 704$\times$525$\times$60 $\mu$m$^3$ (\#2) and 562$\times$225$\times$495 $\mu$m$^3$ (\#3). One of the samples (\#3) was studied with the field perpendicular to the $c$~axis, while the others (\#1  and \#2) were studied with the applied field along the $c$~axis. The coil that generates the ac field produces approximately 2 $\mu$T, which is much smaller than the lower critical field $H_{c1}$, thus ensuring that the sample remains in the Meissner state.

\section{RESULTS}

\begin{figure}[htbp]
\includegraphics[angle=0,width=0.49\textwidth]{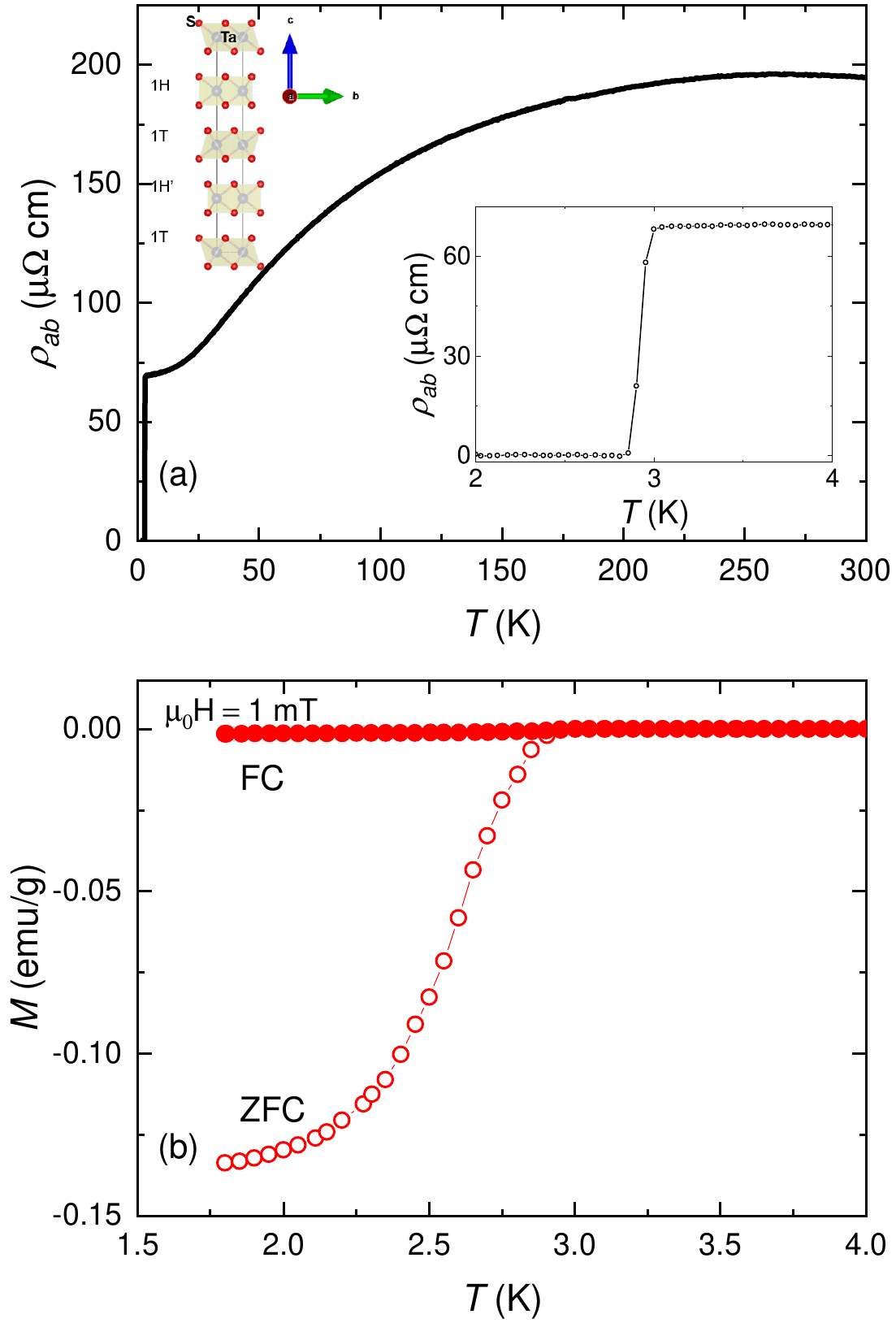}
\vspace{-12pt} \caption{\label{figure1}(Color online) (a) Temperature dependence of the in-plane electrical resistivity $\rho_{ab}(T)$ of 4$H_{b}$-TaS$_{1.99}$Se$_{0.01}$ from room temperature down to 2~K. The top-left and bottom-right insets show the crystal structure and the $\rho_{ab}(T)$ near the superconducting transition, respectively. (b) The temperature dependence of magnetization under both zero-field cooling (ZFC) and field- cooling (FC) processes, with a 10~Oe field applied within the $ab$ plane.}
\vspace{-12pt}
\end{figure}

\begin{figure}[htbp]
\includegraphics[angle=0,width=0.49\textwidth]{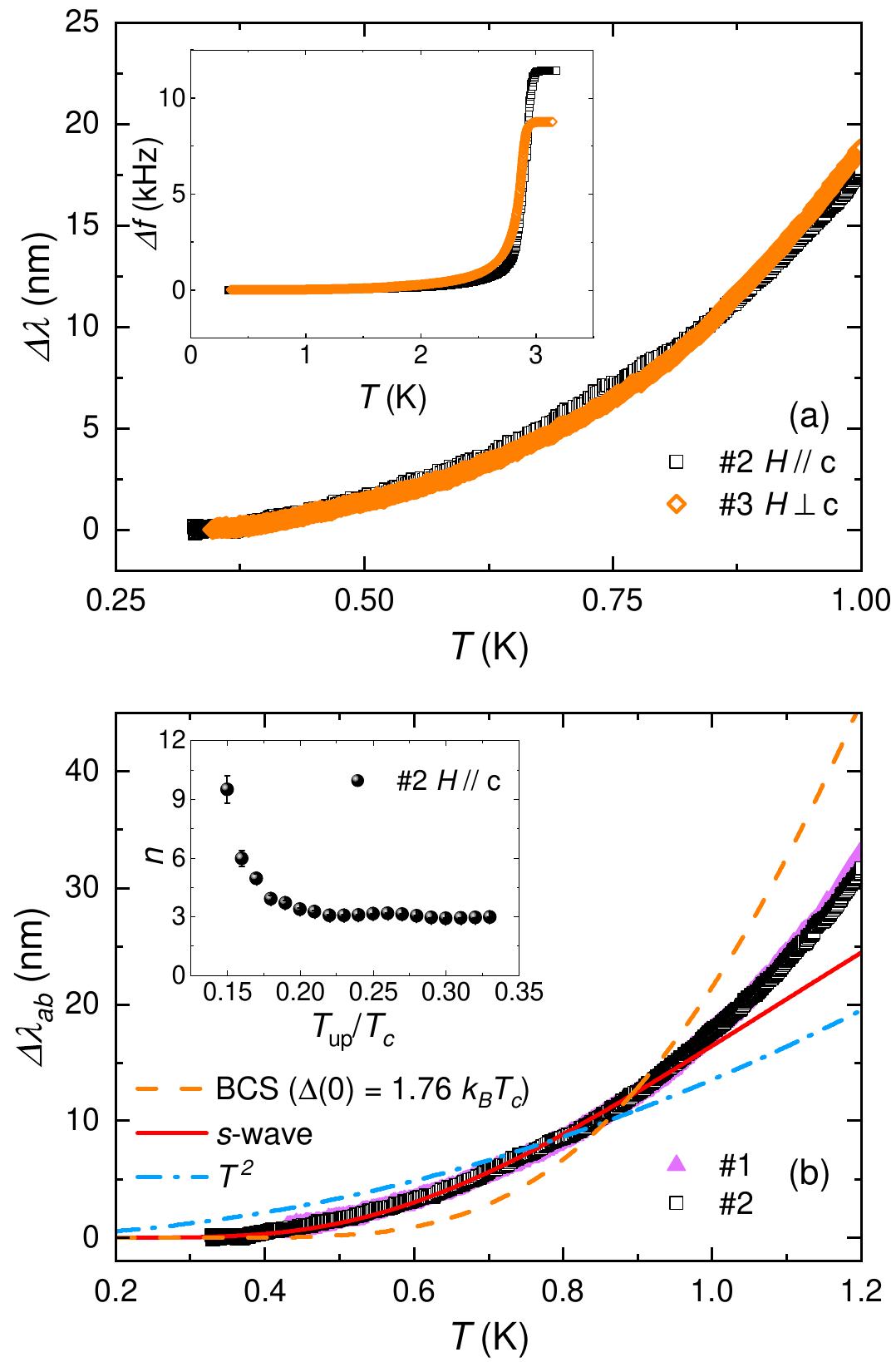}
\vspace{-12pt} \caption{\label{figure2}(Color online) (a) Temperature dependence of the magnetic penetration depth shift $\Delta\lambda(T)$ of 4$H_{b}$-TaS$_{1.99}$Se$_{0.01}$ for applied fields along and perpendicular to the $c$ axis. The inset shows the corresponding $\Delta f(T)$ in the temperature range 3.2 K to 0.33 K. (b) $\Delta\lambda_{ab}(T)$ at low temperatures, measured for two samples with fields along the $c$~axis. The solid, dashed-dotted, and dashed  lines represent fitting to an $s$-wave model, quadratic dependence $\sim T^2$, and the weak coupling BCS model, respectively. The exponent $n$ obtained from a power law fit over different temperature ranges, from the lowest measured temperature up to a temperature $T_{up}$, is shown in the inset.}
\vspace{-12pt}
\end{figure}

The temperature dependence of the $ab$ plane resistivity $\rho_{ab}(T)$ of a 4$H_{b}$-TaS$_{1.99}$Se$_{0.01}$ single crystal from 300~K down to 2~K is displayed in Fig.~\ref{figure1}(a). The $\rho_{ab}(T)$ exhibits metallic behavior with a residual normal state resistivity $\rho_0$ of about $69.3~\mu\Omega$~cm. The inset shows $\rho_{ab}(T)$ at low temperatures, revealing a clear superconducting transition with an onset at $T_{\rm c}^{onset}=2.96$~K and zero resistivity at $T_{\rm c}^{zero}=2.85$~K, consistent with previous results~\cite{meng2024extreme}. The temperature dependence of the magnetization is shown in Fig.~\ref{figure1}(b) under zero-field and field-cooling processes in an applied magnetic field of 10~Oe. A clear superconducting transition is observed with an onset at around 2.96~K. 
Using the values of $\gamma_{n}$~=6.228~mJ mol$^{-1}$~K$^{-2}$ from the specific-heat and the $B_{c2}(0)=1.1$~T~\cite{meng2024extreme}, the penetration depth at zero temperature $\lambda(0)$ is estimated to be 465~nm using $\lambda(0)=\sqrt{\Phi_0B_{c2}(0)}/\sqrt{24\gamma_n}\Delta(0)$~\cite{gross1986anomalous}, consistent with the experimental value $\lambda(0)=487$~nm derived from $\mu\mathrm{SR}$ \cite{ribak2020chiral}. Further combined with $\rho_0= 69.3~\mu\Omega$~cm, the mean free path $\ell$ is estimated to be $\ell$~=74.5~nm~\cite{orlando1979critical}. The BCS coherence length $\xi_{BCS}$ can be calculated using $\xi_{GL}~=~0.74\xi_{BCS}$, where the Ginzburg-Landau (GL) coherence length is given by $\xi_{GL}=\sqrt{\Phi/2\pi B_{c2}(0)}=17.26$~nm, which yields $\xi_{BCS}~=~23.3~nm$. It can be seen that the mean free path is larger than $\xi_{BCS}$, indicating that the sample is in the clean limit.

The temperature dependence of the frequency shift, $\Delta f(T)$, from 3.2~K down to 0.33~K with the applied field both along and perpendicular to the $c$~axis are shown in the inset of Fig.~\ref{figure2}(a), where a distinct reduction upon cooling around $T_\mathrm{c}$~=~2.96~K is observed due to superconductivity, consistent with resistivity, magnetic susceptibility, and specific heat measurements. The low-temperature $\Delta\lambda(T)$ are displayed in Fig.~\ref{figure2}(a), with calibration
factors of \emph{G}~=~13.8~$\textrm{{\AA}/Hz}$ and 10~$\textrm{{\AA}/Hz}$ for the samples with $H\parallel c$ and $H\perp c$, respectively. $\Delta\lambda(T)$ displays nearly identical behavior in both orientations, suggesting that the superconducting state is unlikely to be strongly anisotropic. For $H$~$\parallel$~c, the screening currents flow only in the $ab$ plane, probing the in-plane penetration depth shift $\Delta\lambda_{ab}(T)$~\cite{fletcher2007penetration,prozorov2000meissner}.  $\Delta\lambda_{ab}(T)$ is plotted for two samples in the main panel of Fig.~\ref{figure2}(b) at low temperatures, and this change is derived from $\Delta f(T)$ using \emph{G}~=~15.7~$\textrm{{\AA}/Hz}$ and 13.8~$\textrm{{\AA}/Hz}$ calculated for samples \#1 and \#2, respectively. 
It is clear that the two samples consistently deviate markedly from the $T$-linear and $T^2$ dependences, which are expected for clean-limit superconductors with line nodes and point nodes, respectively. For an isotropic single-band s-wave superconductor, $\Delta\lambda_{ab}(T)$ at $T~\ll~T_\mathrm{c}$ can be approximated by 

\begin{equation}
\Delta\lambda_{ab}(T)=\lambda(0)\sqrt{\frac{\pi\Delta(0)}{2k_\mathrm{B}T}}\textrm{exp}\left(-\frac{\Delta(0)}{k_\mathrm{B}T}\right),
\label{equation1}
\end{equation}

\noindent where $\Delta(0)$ is the superconducting gap magnitude at zero temperature. As shown by the solid red line, the experimental data below $T_\mathrm{c}/3$ are well described by this model with $\Delta(0) = 0.99(2) k_\mathrm{B} T_\mathrm{c}$, while it shows a clear deviation from a model with the gap size of $\Delta(0) = 1.76 k_\mathrm{B} T_\mathrm{c}$ derived from weak-coupling BCS theory. The exponential behavior is further supported by fitting the data to a power law, $\Delta\lambda_{ab}(T)\sim T^n$, where the exponent $n$, obtained from fitting from the lowest measured temperature up to a temperature $T_{up}$ (inset of Fig.~\ref{figure2}(b)), consistently exceeds two and increases as the temperature decreases. The deduced superconducting gap from fitting with Eq.~\ref{equation1} of $0.99 k_\mathrm{B} T_\mathrm{c}$ being smaller than the value for the weak-coupling BCS theory, $1.76 k_\mathrm{B} T_\mathrm{c}$, indicates the presence of multiple gaps and/or gap anisotropy.

To further investigate the gap structure of 4$H_{b}$-TaS$_{1.99}$Se$_{0.01}$, we calculated the superfluid density, $\rho_s(T)$, for samples in the clean limit, using  $\rho_s(T) = [\lambda(0)/\lambda(T)]^2$, with $\lambda(0) = 487$~nm obtained from $\mu\mathrm{SR}$~\cite{ribak2020chiral}. The $\rho_s(T)$ derived from $\mu\mathrm{SR}$ measurements in Ref~\cite{ribak2020chiral} are also displayed, which are consistent with the TDO-method results. The superfluid density is fitted utilizing various gap functions $\Delta_k(T)$, calculated using
\begin{equation}
\rho_{\rm s}(T)=1+2 \left\langle\int_{\Delta_k}^{\infty}\frac{EdE}{\sqrt{E^2-\Delta_k(T)^2}}\frac{\partial f}{\partial E}\right\rangle_{\rm FS},
\label{equation2}
\end{equation}\noindent where $f(E,T)$~=~[1+exp(\emph{E}/$k_\mathrm{B}T)]^{-1}$ is the Fermi-Dirac distribution and $\left\langle...\right\rangle_{\rm FS}$ represents an average over the Fermi surface~\cite{prozorov2006magnetic}. The gap function $\Delta_k$ is defined as $\Delta_k(T)=\Delta(T)g_k$, where $\Delta(T)$ corresponds to the temperature dependence of the gap, approximated as~\cite{carrington2003magnetic} 
\begin{equation}
\Delta(T)~=~\Delta(0){\rm tanh}\left\{1.82\left[1.018\left(T_\mathrm{c}/T-1\right)\right]^{0.51}\right\},
\label{equation3}
\end{equation}
\noindent and $g_k$ represents the angular dependence of the gap structure. Specifically, $g_k$~=~1, sin~$\theta$, and cos~2$\phi$ correspond to $s$-, $p$-, and $d$-wave superconducting gaps, respectively ($\theta$ is the polar angle, and $\phi$ is the azimuthal angle). The data were fitted up to 2.7~K, as above this temperature, the frequency changes increase sharply, as shown in the inset of Fig.~\ref{figure2}(a), which is the onset of the transition to the normal state.

\begin{figure}[t]
\includegraphics[angle=0,width=0.49\textwidth]{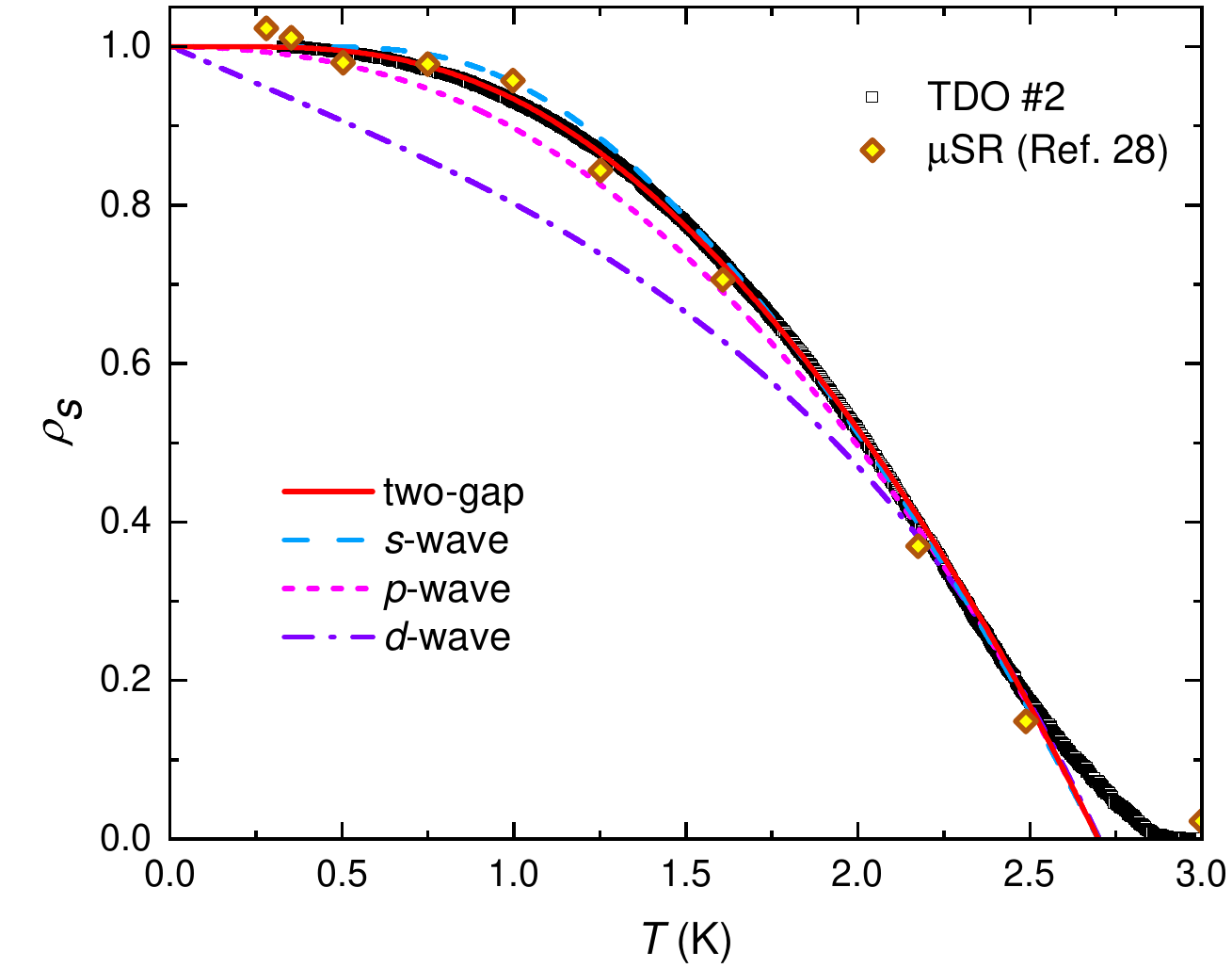}
\vspace{-12pt} \caption{\label{figure3}(Color online) Temperature dependence of the normalized superfluid density $\rho_s(T)$ for sample \#2 with $\lambda(0)=487$~nm. The solid, dashed, short-dashed, and dashed-dotted lines represent the fitting results with two-gap $s$-wave, $s$-wave, $p$-wave, $d$-wave models, respectively. The  rhombus symbols show results derived from $\mu\mathrm{SR}$ measurements from Ref.~\cite{ribak2020chiral}.}
\vspace{-12pt}
\end{figure}
Consistent with the analysis of the magnetic penetration depth, the superfluid density exhibits significant deviations from both the $p$-wave and $d$-wave models (Fig.~\ref{figure3}), which have point nodes and line nodes, respectively. A single gap $s$-wave model still does not describe the data well at low temperatures. In contrast, the data across the entire temperature range is fitted well by a two-gap $s$-wave model~\cite{carrington2003magnetic}, described by $\rho_{\rm s}(T)=x\rho_{\rm s1}+(1-x)\rho_{\rm s2} $, where $\rho_{\rm s1}$ and $\rho_{\rm s2}$ are calculated using Eq.~\ref{equation2} for gap magnitudes $\Delta_1(0)$ and $\Delta_2(0)$, respectively, and $x$ is the weight of the contribution from $\Delta_1(0)$. The fitted parameters are $\Delta_1(0)=0.89(2)k_\mathrm{B}T_\mathrm{c}$, $\Delta_2(0)=1.96(1)k_\mathrm{B}T_\mathrm{c}$ and $x=0.099(6)$, which corresponds to a weight of the smaller gap of around 10$\%$. The small gap obtained from fitting the superfluid density with a two-gap $s$-wave model is similar to the fitted gap from the analysis of the low-temperature $\Delta\lambda_{ab}$, further corroborating the presence of multiple superconducting gaps.

\begin{figure}[t]
\includegraphics[angle=0,width=0.49\textwidth]{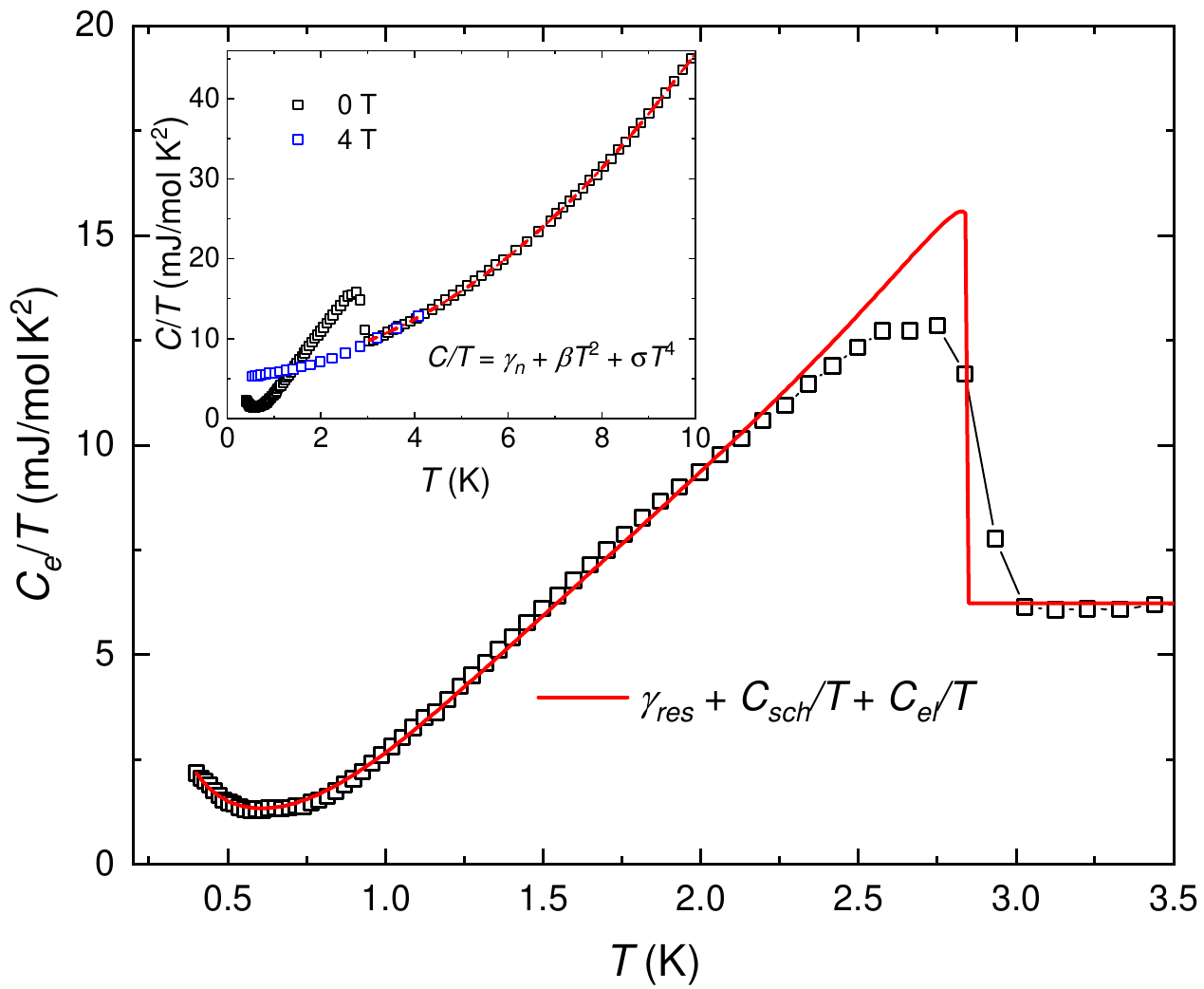}
\vspace{-12pt} \caption{\label{figure4}(Color online) Temperature dependence of the electronic specific heat as $C_e/T$ of 4$H_{b}$-TaS$_{1.99}$Se$_{0.01}$, where the solid line represents the fit with a two-gap $s$-wave model, including a Schottky contribution. The inset shows the total specific heat $C(T)/T$ in zero-field and 4~T, where the dashed line represents the fitting to the normal state contribution.}
\vspace{-12pt}
\end{figure}

The total specific heat, plotted as $C(T)/T$, is displayed in the inset of Fig.~\ref{figure4}, showing a clear superconducting transition at around $T_\mathrm{c} = 2.85~\mathrm{K}$, providing evidence for bulk superconductivity. In the normal state, the data were fitted using $C(T)/T = \gamma_n +\beta T^2 + \delta T^4$, where $\gamma_n = 6.228~\mathrm{mJ~mol^{-1}~K^{-2}}$, $\beta = 0.386~\mathrm{mJ~mol^{-1}~K^{-4}}$, and $\delta = 0.087~\mu\mathrm{J~mol^{-1}~K^{-6}}$. Here, $C_e = \gamma_n T$ and $C_{ph} = \beta T^3 + \delta T^5$ represent the contributions from normal-state electrons and phonons, respectively, with $\gamma_n$ being the Sommerfeld coefficient. At low temperatures, $C(T)/T$ shows a noticeable upturn which is suppressed in a high field, likely due to a nuclear Schottky anomaly. After subtracting the phonon contributions, the electronic specific heat $C_{e}(T)$ of the superconducting state can be described by~\cite{wang2001specific,yang2001order}:
\begin{equation}
C_e(T) = \gamma_{res}T + C_\mathrm{Schottky}(\Delta E/k_\mathrm{B}T) + TdS_{sc}/dT,
\label{equation4}
\end{equation}
\noindent where $\gamma_{res}$ represents the residual contribution. The second term is a two-level Schottky anomaly, given by $C_\mathrm{Schottky}(x) = \frac{nRx^2 e^x}{(1+e^x)^2}$, with $\Delta E$ being the level splitting energy, \emph{R} the gas constant and \emph{n} the concentration of paramagnetic impurities. The entropy $S_{sc}$ of a single gap superconductor can be calculated by~\cite{bouquet2001phenomenological}
\begin{equation}
  S_{sc}~=~-\frac{3\gamma_n}{\pi^3}\int_0^{2\pi}\int_0^\infty[f\textrm{ln}f+(1-f)\textrm{ln}(1-f)]d\varepsilon  d\phi.
 \label{equation5}
 \end{equation}
\noindent We analyzed $C_e(T)$ using Eq.~\ref{equation4}, where $dS_{sc}/dT$ is calculated using a two gap model, which consists of the sum of single band contributions with gap magnitudes $\Delta_1$ and $\Delta_2$ (Eq.~\ref{equation5}), with a relative weight $x$ of the contribution from the former gap. Here we fixed the values of the parameters $\Delta_1(0)$, $\Delta_2(0)$ and $x$ to those from the analysis of the superfluid density, so that the fitted parameters are the two associated with the Schottky term $\Delta E$ and \emph{n}, as well as $\gamma_{res}$. As shown by the solid line in Fig.~\ref{figure4}, $C_{e}(T)$ is well fitted throughout the superconducting state, with parameters $\Delta E = 4.6 \times 10^{-24}~\mathrm{J}$, $n = 0.00058$, and $\gamma_{res} = 0.3~\mathrm{mJ~mol^{-1}~K^{-2}}$. The residual contribution  $\gamma_{res}$ is similar to previous results~\cite{ribak2020chiral}, which is likely due to a nonsuperconducting  fraction, possibly originating from different stacking arrangements of 1$T$ and 1$H$ layers in 4$H_{b}$-TaS$_{1.99}$Se$_{0.01}$. 
\section{\label{sec:citeref}Discussion and conclusions}
The exponentially-activated behavior of $\Delta\lambda(T)$ at low temperatures evidences nodeless superconductivity in 4$H_{b}$-TaS$_{1.99}$Se$_{0.01}$, consistent with previous studies of the specific heat and $\mu\mathrm{SR}$~\cite{ribak2020chiral}. Moreover, our analysis of $\Delta\lambda(T)$ and the derived superfluid support the presence of multigap superconductivity, whereby the gap magnitude from the analysis of $\Delta\lambda(T)$ at low temperatures is considerably less than the weak coupling BCS value, and is also much smaller than that from fitting the superfluid density to a model with a single isotropic gap. Furthermore this one-gap model cannot well describe the superfluid density over the full temperature range. In contrast, the superfluid density is well fitted by a two-gap $s$-wave model, and the smaller gap from this fitting is close to the value from the low temperature analysis of $\Delta\lambda(T)$, providing clear evidence for multigap superconductivity. Moreover, the same parameters deduced from fitting the two-gap model to the superfluid density, can also consistently describe the specific heat.  

The conclusion of multigap superconductivity is consistent with presence of multiple bands crossing the Fermi energy, as revealed by angle-resolved photoemission spectroscopy (ARPES) measurements and electronic structure calculations~\cite{almoalem2024charge}. Besides the inner pocket along the $K-M-K$' line, which originates from the 1$H$ layer, ARPES also observes several shallow pockets around the $\Gamma$ point in the normal state, which originate from the 1$T$ layers but are notably different from the bulk bands of 1$T$-TaS$_2$. This is further supported by STM results, which identify depleted flat bands with a weak density of states crossing the Fermi level~\cite{kumar2023first}. These different bands may play an important role in the superconductivity and are likely associated with superconducting gaps of different sizes.

Based on the analysis of the irreducible representations (irreps) of the hexagonal point group and the pronounced two-dimensional nature of 4$H_{b}$-TaS$_{1.99}$Se$_{0.01}$, only four irreps need to be considered: the trivial irrep \(A_{1g}\), \(B_{1u}\), and the two-dimensional irreps \(E_{2g}\) and \(E_{1u}\)~\cite{fischer2015symmetry,almoalem2024observation}. Our results demonstrate that the superconducting gap structure of 4$H_{b}$-TaS$_{1.99}$Se$_{0.01}$ is nodeless, ruling out the \(B_{1u}\) and \(E_{2g}\) order parameters, as both exhibit symmetry enforced nodes in the gap function. Considering the TRSB inferred from $\mu\mathrm{SR}$ measurements~\cite{ribak2020chiral}, complex \(p+ip\) states are favored over the even-parity \(A_{1g}\) representation, but this identification requires further verification via phase-sensitive experiments.

In summary, we have investigated the superconducting pairing symmetry of 4$H_{b}$-TaS$_{1.99}$Se$_{0.01}$ single crystals using magnetic penetration depth measurements and the specific heat. The penetration depth data below $T_\mathrm{c}$/3 show a clear exponentially activated temperature dependence, indicating a nodeless superconducting gap. The temperature dependence of the superfluid density and specific heat can be well described by nodeless two-gap superconductivity, which is consistent with the presence of multiple Fermi surfaces originating from both the 1$H$ and 1$T$ layers in 4$H_{b}$-TaS$_{1.99}$Se$_{0.01}$.

\begin{acknowledgments}
 This work was supported by the National Key R\&D Program of China (Grant No. 2022YFA1402200, No.
2023YFA1406303, No. 2022YFA1403800 and 2023YFA1406500),  the National Natural Science Foundation of China (Grants No. 12034017, No. 12494592, No. 12222410, No. 12274459 and No. 12174332)
\end{acknowledgments}
%

\nocite{*}


\begin{thebibliography}{48}%
\makeatletter
\providecommand \@ifxundefined [1]{%
 \@ifx{#1\undefined}
}%
\providecommand \@ifnum [1]{%
 \ifnum #1\expandafter \@firstoftwo
 \else \expandafter \@secondoftwo
 \fi
}%
\providecommand \@ifx [1]{%
 \ifx #1\expandafter \@firstoftwo
 \else \expandafter \@secondoftwo
 \fi
}%
\providecommand \natexlab [1]{#1}%
\providecommand \enquote  [1]{``#1''}%
\providecommand \bibnamefont  [1]{#1}%
\providecommand \bibfnamefont [1]{#1}%
\providecommand \citenamefont [1]{#1}%
\providecommand \href@noop [0]{\@secondoftwo}%
\providecommand \href [0]{\begingroup \@sanitize@url \@href}%
\providecommand \@href[1]{\@@startlink{#1}\@@href}%
\providecommand \@@href[1]{\endgroup#1\@@endlink}%
\providecommand \@sanitize@url [0]{\catcode `\\12\catcode `\$12\catcode
  `\&12\catcode `\#12\catcode `\^12\catcode `\_12\catcode `\%12\relax}%
\providecommand \@@startlink[1]{}%
\providecommand \@@endlink[0]{}%
\providecommand \url  [0]{\begingroup\@sanitize@url \@url }%
\providecommand \@url [1]{\endgroup\@href {#1}{\urlprefix }}%
\providecommand \urlprefix  [0]{URL }%
\providecommand \Eprint [0]{\href }%
\providecommand \doibase [0]{https://doi.org/}%
\providecommand \selectlanguage [0]{\@gobble}%
\providecommand \bibinfo  [0]{\@secondoftwo}%
\providecommand \bibfield  [0]{\@secondoftwo}%
\providecommand \translation [1]{[#1]}%
\providecommand \BibitemOpen [0]{}%
\providecommand \bibitemStop [0]{}%
\providecommand \bibitemNoStop [0]{.\EOS\space}%
\providecommand \EOS [0]{\spacefactor3000\relax}%
\providecommand \BibitemShut  [1]{\csname bibitem#1\endcsname}%
\let\auto@bib@innerbib\@empty
\bibitem [{\citenamefont {Sigrist}\ and\ \citenamefont
  {Ueda}(1991)}]{sigrist1991phenomenological}%
  \BibitemOpen
  \bibfield  {author} {\bibinfo {author} {\bibfnamefont {M.}~\bibnamefont
  {Sigrist}}\ and\ \bibinfo {author} {\bibfnamefont {K.}~\bibnamefont {Ueda}},\
  }\bibfield  {title} {\bibinfo {title} {Phenomenological theory of
  unconventional superconductivity},\ }\href@noop {} {\bibfield  {journal}
  {\bibinfo  {journal} {Reviews of Modern physics}\ }\textbf {\bibinfo {volume}
  {63}},\ \bibinfo {pages} {239} (\bibinfo {year} {1991})}\BibitemShut
  {NoStop}%
\bibitem [{\citenamefont {Ghosh}\ \emph {et~al.}(2020)\citenamefont {Ghosh},
  \citenamefont {Smidman}, \citenamefont {Shang}, \citenamefont {Annett},
  \citenamefont {Hillier}, \citenamefont {Quintanilla},\ and\ \citenamefont
  {Yuan}}]{ghosh2020recent}%
  \BibitemOpen
  \bibfield  {author} {\bibinfo {author} {\bibfnamefont {S.~K.}\ \bibnamefont
  {Ghosh}}, \bibinfo {author} {\bibfnamefont {M.}~\bibnamefont {Smidman}},
  \bibinfo {author} {\bibfnamefont {T.}~\bibnamefont {Shang}}, \bibinfo
  {author} {\bibfnamefont {J.~F.}\ \bibnamefont {Annett}}, \bibinfo {author}
  {\bibfnamefont {A.~D.}\ \bibnamefont {Hillier}}, \bibinfo {author}
  {\bibfnamefont {J.}~\bibnamefont {Quintanilla}},\ and\ \bibinfo {author}
  {\bibfnamefont {H.}~\bibnamefont {Yuan}},\ }\bibfield  {title} {\bibinfo
  {title} {Recent progress on superconductors with time-reversal symmetry
  breaking},\ }\href@noop {} {\bibfield  {journal} {\bibinfo  {journal}
  {Journal of Physics: Condensed Matter}\ }\textbf {\bibinfo {volume} {33}},\
  \bibinfo {pages} {033001} (\bibinfo {year} {2020})}\BibitemShut {NoStop}%
\bibitem [{\citenamefont {Luke}\ \emph {et~al.}(1993)\citenamefont {Luke},
  \citenamefont {Keren}, \citenamefont {Le}, \citenamefont {Wu}, \citenamefont
  {Uemura}, \citenamefont {Bonn}, \citenamefont {Taillefer},\ and\
  \citenamefont {Garrett}}]{luke1993muon}%
  \BibitemOpen
  \bibfield  {author} {\bibinfo {author} {\bibfnamefont {G.}~\bibnamefont
  {Luke}}, \bibinfo {author} {\bibfnamefont {A.}~\bibnamefont {Keren}},
  \bibinfo {author} {\bibfnamefont {L.}~\bibnamefont {Le}}, \bibinfo {author}
  {\bibfnamefont {W.}~\bibnamefont {Wu}}, \bibinfo {author} {\bibfnamefont
  {Y.}~\bibnamefont {Uemura}}, \bibinfo {author} {\bibfnamefont
  {D.}~\bibnamefont {Bonn}}, \bibinfo {author} {\bibfnamefont {L.}~\bibnamefont
  {Taillefer}},\ and\ \bibinfo {author} {\bibfnamefont {J.}~\bibnamefont
  {Garrett}},\ }\bibfield  {title} {\bibinfo {title} {Muon spin relaxation in
  {U}{Pt}$_3$},\ }\href@noop {} {\bibfield  {journal} {\bibinfo  {journal}
  {Physical review letters}\ }\textbf {\bibinfo {volume} {71}},\ \bibinfo
  {pages} {1466} (\bibinfo {year} {1993})}\BibitemShut {NoStop}%
\bibitem [{\citenamefont {Luke}\ \emph {et~al.}(1998)\citenamefont {Luke},
  \citenamefont {Fudamoto}, \citenamefont {Kojima}, \citenamefont {Larkin},
  \citenamefont {Merrin}, \citenamefont {Nachumi}, \citenamefont {Uemura},
  \citenamefont {Maeno}, \citenamefont {Mao}, \citenamefont {Mori} \emph
  {et~al.}}]{luke1998time}%
  \BibitemOpen
  \bibfield  {author} {\bibinfo {author} {\bibfnamefont {G.~M.}\ \bibnamefont
  {Luke}}, \bibinfo {author} {\bibfnamefont {Y.}~\bibnamefont {Fudamoto}},
  \bibinfo {author} {\bibfnamefont {K.}~\bibnamefont {Kojima}}, \bibinfo
  {author} {\bibfnamefont {M.}~\bibnamefont {Larkin}}, \bibinfo {author}
  {\bibfnamefont {J.}~\bibnamefont {Merrin}}, \bibinfo {author} {\bibfnamefont
  {B.}~\bibnamefont {Nachumi}}, \bibinfo {author} {\bibfnamefont
  {Y.}~\bibnamefont {Uemura}}, \bibinfo {author} {\bibfnamefont
  {Y.}~\bibnamefont {Maeno}}, \bibinfo {author} {\bibfnamefont
  {Z.}~\bibnamefont {Mao}}, \bibinfo {author} {\bibfnamefont {Y.}~\bibnamefont
  {Mori}}, \emph {et~al.},\ }\bibfield  {title} {\bibinfo {title}
  {Time-reversal symmetry-breaking superconductivity in {Sr}$_2${Ru}{O}$_4$},\
  }\href@noop {} {\bibfield  {journal} {\bibinfo  {journal} {Nature}\ }\textbf
  {\bibinfo {volume} {394}},\ \bibinfo {pages} {558} (\bibinfo {year}
  {1998})}\BibitemShut {NoStop}%
\bibitem [{\citenamefont {Xia}\ \emph {et~al.}(2006)\citenamefont {Xia},
  \citenamefont {Maeno}, \citenamefont {Beyersdorf}, \citenamefont {Fejer},\
  and\ \citenamefont {Kapitulnik}}]{xia2006high}%
  \BibitemOpen
  \bibfield  {author} {\bibinfo {author} {\bibfnamefont {J.}~\bibnamefont
  {Xia}}, \bibinfo {author} {\bibfnamefont {Y.}~\bibnamefont {Maeno}}, \bibinfo
  {author} {\bibfnamefont {P.~T.}\ \bibnamefont {Beyersdorf}}, \bibinfo
  {author} {\bibfnamefont {M.}~\bibnamefont {Fejer}},\ and\ \bibinfo {author}
  {\bibfnamefont {A.}~\bibnamefont {Kapitulnik}},\ }\bibfield  {title}
  {\bibinfo {title} {High resolution polar kerr effect measurements of
  {Sr}$_2${Ru}{O}$_4$: Evidence for broken time-reversal symmetry in the
  superconducting state},\ }\href@noop {} {\bibfield  {journal} {\bibinfo
  {journal} {Physical review letters}\ }\textbf {\bibinfo {volume} {97}},\
  \bibinfo {pages} {167002} (\bibinfo {year} {2006})}\BibitemShut {NoStop}%
\bibitem [{\citenamefont {Heffner}\ \emph {et~al.}(1990)\citenamefont
  {Heffner}, \citenamefont {Smith}, \citenamefont {Willis}, \citenamefont
  {Birrer}, \citenamefont {Baines}, \citenamefont {Gygax}, \citenamefont
  {Hitti}, \citenamefont {Lippelt}, \citenamefont {Ott}, \citenamefont
  {Schenck} \emph {et~al.}}]{heffner1990new}%
  \BibitemOpen
  \bibfield  {author} {\bibinfo {author} {\bibfnamefont {R.}~\bibnamefont
  {Heffner}}, \bibinfo {author} {\bibfnamefont {J.}~\bibnamefont {Smith}},
  \bibinfo {author} {\bibfnamefont {J.}~\bibnamefont {Willis}}, \bibinfo
  {author} {\bibfnamefont {P.}~\bibnamefont {Birrer}}, \bibinfo {author}
  {\bibfnamefont {C.}~\bibnamefont {Baines}}, \bibinfo {author} {\bibfnamefont
  {F.}~\bibnamefont {Gygax}}, \bibinfo {author} {\bibfnamefont
  {B.}~\bibnamefont {Hitti}}, \bibinfo {author} {\bibfnamefont
  {E.}~\bibnamefont {Lippelt}}, \bibinfo {author} {\bibfnamefont
  {H.}~\bibnamefont {Ott}}, \bibinfo {author} {\bibfnamefont {A.}~\bibnamefont
  {Schenck}}, \emph {et~al.},\ }\bibfield  {title} {\bibinfo {title} {New phase
  diagram for ({U},{Th}){Be}$_{13}$: A muon-spin-resonance and
  $\textit{H}_{c1}$ study},\ }\href@noop {} {\bibfield  {journal} {\bibinfo
  {journal} {Physical review letters}\ }\textbf {\bibinfo {volume} {65}},\
  \bibinfo {pages} {2816} (\bibinfo {year} {1990})}\BibitemShut {NoStop}%
\bibitem [{\citenamefont {Schemm}\ \emph {et~al.}(2015)\citenamefont {Schemm},
  \citenamefont {Baumbach}, \citenamefont {Tobash}, \citenamefont {Ronning},
  \citenamefont {Bauer},\ and\ \citenamefont
  {Kapitulnik}}]{schemm2015evidence}%
  \BibitemOpen
  \bibfield  {author} {\bibinfo {author} {\bibfnamefont {E.}~\bibnamefont
  {Schemm}}, \bibinfo {author} {\bibfnamefont {R.}~\bibnamefont {Baumbach}},
  \bibinfo {author} {\bibfnamefont {P.}~\bibnamefont {Tobash}}, \bibinfo
  {author} {\bibfnamefont {F.}~\bibnamefont {Ronning}}, \bibinfo {author}
  {\bibfnamefont {E.}~\bibnamefont {Bauer}},\ and\ \bibinfo {author}
  {\bibfnamefont {A.}~\bibnamefont {Kapitulnik}},\ }\bibfield  {title}
  {\bibinfo {title} {Evidence for broken time-reversal symmetry in the
  superconducting phase of {U}{Ru}$_2${Si}$_2$},\ }\href@noop {} {\bibfield
  {journal} {\bibinfo  {journal} {Physical Review B}\ }\textbf {\bibinfo
  {volume} {91}},\ \bibinfo {pages} {140506} (\bibinfo {year}
  {2015})}\BibitemShut {NoStop}%
\bibitem [{\citenamefont {Schemm}\ \emph {et~al.}(2014)\citenamefont {Schemm},
  \citenamefont {Gannon}, \citenamefont {Wishne}, \citenamefont {Halperin},\
  and\ \citenamefont {Kapitulnik}}]{schemm2014observation}%
  \BibitemOpen
  \bibfield  {author} {\bibinfo {author} {\bibfnamefont {E.}~\bibnamefont
  {Schemm}}, \bibinfo {author} {\bibfnamefont {W.}~\bibnamefont {Gannon}},
  \bibinfo {author} {\bibfnamefont {C.}~\bibnamefont {Wishne}}, \bibinfo
  {author} {\bibfnamefont {W.}~\bibnamefont {Halperin}},\ and\ \bibinfo
  {author} {\bibfnamefont {A.}~\bibnamefont {Kapitulnik}},\ }\bibfield  {title}
  {\bibinfo {title} {Observation of broken time-reversal symmetry in the
  heavy-fermion superconductor {U}{Pt}$_3$},\ }\href@noop {} {\bibfield
  {journal} {\bibinfo  {journal} {Science}\ }\textbf {\bibinfo {volume}
  {345}},\ \bibinfo {pages} {190} (\bibinfo {year} {2014})}\BibitemShut
  {NoStop}%
\bibitem [{\citenamefont {Hillier}\ \emph {et~al.}(2009)\citenamefont
  {Hillier}, \citenamefont {Quintanilla},\ and\ \citenamefont
  {Cywinski}}]{hillier2009evidence}%
  \BibitemOpen
  \bibfield  {author} {\bibinfo {author} {\bibfnamefont {A.~D.}\ \bibnamefont
  {Hillier}}, \bibinfo {author} {\bibfnamefont {J.}~\bibnamefont
  {Quintanilla}},\ and\ \bibinfo {author} {\bibfnamefont {R.}~\bibnamefont
  {Cywinski}},\ }\bibfield  {title} {\bibinfo {title} {Evidence for
  time-reversal symmetry breaking in the noncentrosymmetric superconductor
  {La}{Ni}{C}$_2$},\ }\href@noop {} {\bibfield  {journal} {\bibinfo  {journal}
  {Physical review letters}\ }\textbf {\bibinfo {volume} {102}},\ \bibinfo
  {pages} {117007} (\bibinfo {year} {2009})}\BibitemShut {NoStop}%
\bibitem [{\citenamefont {Hillier}\ \emph {et~al.}(2012)\citenamefont
  {Hillier}, \citenamefont {Quintanilla}, \citenamefont {Mazidian},
  \citenamefont {Annett},\ and\ \citenamefont
  {Cywinski}}]{hillier2012nonunitary}%
  \BibitemOpen
  \bibfield  {author} {\bibinfo {author} {\bibfnamefont {A.~D.}\ \bibnamefont
  {Hillier}}, \bibinfo {author} {\bibfnamefont {J.}~\bibnamefont
  {Quintanilla}}, \bibinfo {author} {\bibfnamefont {B.}~\bibnamefont
  {Mazidian}}, \bibinfo {author} {\bibfnamefont {J.~F.}\ \bibnamefont
  {Annett}},\ and\ \bibinfo {author} {\bibfnamefont {R.}~\bibnamefont
  {Cywinski}},\ }\bibfield  {title} {\bibinfo {title} {Nonunitary triplet
  pairing in the centrosymmetric superconductor {La}{Ni}{Ga}$_2$},\ }\href@noop
  {} {\bibfield  {journal} {\bibinfo  {journal} {Physical Review Letters}\
  }\textbf {\bibinfo {volume} {109}},\ \bibinfo {pages} {097001} (\bibinfo
  {year} {2012})}\BibitemShut {NoStop}%
\bibitem [{\citenamefont {Weng}\ \emph {et~al.}(2016)\citenamefont {Weng},
  \citenamefont {Zhang}, \citenamefont {Smidman}, \citenamefont {Shang},
  \citenamefont {Quintanilla}, \citenamefont {Annett}, \citenamefont {Nicklas},
  \citenamefont {Pang}, \citenamefont {Jiao}, \citenamefont {Jiang} \emph
  {et~al.}}]{weng2016two}%
  \BibitemOpen
  \bibfield  {author} {\bibinfo {author} {\bibfnamefont {Z.}~\bibnamefont
  {Weng}}, \bibinfo {author} {\bibfnamefont {J.}~\bibnamefont {Zhang}},
  \bibinfo {author} {\bibfnamefont {M.}~\bibnamefont {Smidman}}, \bibinfo
  {author} {\bibfnamefont {T.}~\bibnamefont {Shang}}, \bibinfo {author}
  {\bibfnamefont {J.}~\bibnamefont {Quintanilla}}, \bibinfo {author}
  {\bibfnamefont {J.~F.}\ \bibnamefont {Annett}}, \bibinfo {author}
  {\bibfnamefont {M.}~\bibnamefont {Nicklas}}, \bibinfo {author} {\bibfnamefont
  {G.}~\bibnamefont {Pang}}, \bibinfo {author} {\bibfnamefont {L.}~\bibnamefont
  {Jiao}}, \bibinfo {author} {\bibfnamefont {W.}~\bibnamefont {Jiang}}, \emph
  {et~al.},\ }\bibfield  {title} {\bibinfo {title} {Two-gap superconductivity
  in {La}{Ni}{Ga}$_2$ with nonunitary triplet pairing and even parity gap
  symmetry},\ }\href@noop {} {\bibfield  {journal} {\bibinfo  {journal}
  {Physical Review Letters}\ }\textbf {\bibinfo {volume} {117}},\ \bibinfo
  {pages} {027001} (\bibinfo {year} {2016})}\BibitemShut {NoStop}%
\bibitem [{\citenamefont {Bhattacharyya}\ \emph {et~al.}(2015)\citenamefont
  {Bhattacharyya}, \citenamefont {Adroja}, \citenamefont {Quintanilla},
  \citenamefont {Hillier}, \citenamefont {Kase}, \citenamefont {Strydom},\ and\
  \citenamefont {Akimitsu}}]{bhattacharyya2015broken}%
  \BibitemOpen
  \bibfield  {author} {\bibinfo {author} {\bibfnamefont {A.}~\bibnamefont
  {Bhattacharyya}}, \bibinfo {author} {\bibfnamefont {D.}~\bibnamefont
  {Adroja}}, \bibinfo {author} {\bibfnamefont {J.}~\bibnamefont {Quintanilla}},
  \bibinfo {author} {\bibfnamefont {A.}~\bibnamefont {Hillier}}, \bibinfo
  {author} {\bibfnamefont {N.}~\bibnamefont {Kase}}, \bibinfo {author}
  {\bibfnamefont {A.}~\bibnamefont {Strydom}},\ and\ \bibinfo {author}
  {\bibfnamefont {J.}~\bibnamefont {Akimitsu}},\ }\bibfield  {title} {\bibinfo
  {title} {Broken time-reversal symmetry probed by muon spin relaxation in the
  caged type superconductor {Lu}$_5${Rh}$_6${Sn}$_{18}$},\ }\href@noop {}
  {\bibfield  {journal} {\bibinfo  {journal} {Physical Review B}\ }\textbf
  {\bibinfo {volume} {91}},\ \bibinfo {pages} {060503} (\bibinfo {year}
  {2015})}\BibitemShut {NoStop}%
\bibitem [{\citenamefont {Wang}\ \emph {et~al.}(2021)\citenamefont {Wang},
  \citenamefont {Nie}, \citenamefont {Du}, \citenamefont {Pang}, \citenamefont
  {Kase}, \citenamefont {Akimitsu}, \citenamefont {Chen}, \citenamefont
  {Gutmann}, \citenamefont {Adroja}, \citenamefont {Perry} \emph
  {et~al.}}]{wang2021nodeless}%
  \BibitemOpen
  \bibfield  {author} {\bibinfo {author} {\bibfnamefont {A.}~\bibnamefont
  {Wang}}, \bibinfo {author} {\bibfnamefont {Z.}~\bibnamefont {Nie}}, \bibinfo
  {author} {\bibfnamefont {F.}~\bibnamefont {Du}}, \bibinfo {author}
  {\bibfnamefont {G.}~\bibnamefont {Pang}}, \bibinfo {author} {\bibfnamefont
  {N.}~\bibnamefont {Kase}}, \bibinfo {author} {\bibfnamefont {J.}~\bibnamefont
  {Akimitsu}}, \bibinfo {author} {\bibfnamefont {Y.}~\bibnamefont {Chen}},
  \bibinfo {author} {\bibfnamefont {M.}~\bibnamefont {Gutmann}}, \bibinfo
  {author} {\bibfnamefont {D.}~\bibnamefont {Adroja}}, \bibinfo {author}
  {\bibfnamefont {R.}~\bibnamefont {Perry}}, \emph {et~al.},\ }\bibfield
  {title} {\bibinfo {title} {Nodeless superconductivity in
  {Lu}$_{5-x}${Rh}$_6${Sn}$_{18+x}$ with broken time reversal symmetry},\
  }\href@noop {} {\bibfield  {journal} {\bibinfo  {journal} {Physical Review
  B}\ }\textbf {\bibinfo {volume} {103}},\ \bibinfo {pages} {024503} (\bibinfo
  {year} {2021})}\BibitemShut {NoStop}%
\bibitem [{\citenamefont {Xie}\ \emph {et~al.}(2020)\citenamefont {Xie},
  \citenamefont {Zhang}, \citenamefont {Shen}, \citenamefont {Jiang},
  \citenamefont {Pang}, \citenamefont {Shang}, \citenamefont {Cao},
  \citenamefont {Smidman},\ and\ \citenamefont {Yuan}}]{xie2020captas}%
  \BibitemOpen
  \bibfield  {author} {\bibinfo {author} {\bibfnamefont {W.}~\bibnamefont
  {Xie}}, \bibinfo {author} {\bibfnamefont {P.}~\bibnamefont {Zhang}}, \bibinfo
  {author} {\bibfnamefont {B.}~\bibnamefont {Shen}}, \bibinfo {author}
  {\bibfnamefont {W.}~\bibnamefont {Jiang}}, \bibinfo {author} {\bibfnamefont
  {G.}~\bibnamefont {Pang}}, \bibinfo {author} {\bibfnamefont {T.}~\bibnamefont
  {Shang}}, \bibinfo {author} {\bibfnamefont {C.}~\bibnamefont {Cao}}, \bibinfo
  {author} {\bibfnamefont {M.}~\bibnamefont {Smidman}},\ and\ \bibinfo {author}
  {\bibfnamefont {H.}~\bibnamefont {Yuan}},\ }\bibfield  {title} {\bibinfo
  {title} {{Ca}{Pt}{As}: A new noncentrosymmetric superconductor},\ }\href@noop
  {} {\bibfield  {journal} {\bibinfo  {journal} {Science China Physics,
  Mechanics \& Astronomy}\ }\textbf {\bibinfo {volume} {63}},\ \bibinfo {pages}
  {237412} (\bibinfo {year} {2020})}\BibitemShut {NoStop}%
\bibitem [{\citenamefont {Shang}\ \emph {et~al.}(2020)\citenamefont {Shang},
  \citenamefont {Smidman}, \citenamefont {Wang}, \citenamefont {Chang},
  \citenamefont {Baines}, \citenamefont {Lee}, \citenamefont {Nie},
  \citenamefont {Pang}, \citenamefont {Xie}, \citenamefont {Jiang} \emph
  {et~al.}}]{shang2020simultaneous}%
  \BibitemOpen
  \bibfield  {author} {\bibinfo {author} {\bibfnamefont {T.}~\bibnamefont
  {Shang}}, \bibinfo {author} {\bibfnamefont {M.}~\bibnamefont {Smidman}},
  \bibinfo {author} {\bibfnamefont {A.}~\bibnamefont {Wang}}, \bibinfo {author}
  {\bibfnamefont {L.-J.}\ \bibnamefont {Chang}}, \bibinfo {author}
  {\bibfnamefont {C.}~\bibnamefont {Baines}}, \bibinfo {author} {\bibfnamefont
  {M.-K.}\ \bibnamefont {Lee}}, \bibinfo {author} {\bibfnamefont
  {Z.}~\bibnamefont {Nie}}, \bibinfo {author} {\bibfnamefont {G.}~\bibnamefont
  {Pang}}, \bibinfo {author} {\bibfnamefont {W.}~\bibnamefont {Xie}}, \bibinfo
  {author} {\bibfnamefont {W.}~\bibnamefont {Jiang}}, \emph {et~al.},\
  }\bibfield  {title} {\bibinfo {title} {Simultaneous nodal superconductivity
  and time-reversal symmetry breaking in the noncentrosymmetric superconductor
  {Ca}{Pt}{As}},\ }\href@noop {} {\bibfield  {journal} {\bibinfo  {journal}
  {Physical review letters}\ }\textbf {\bibinfo {volume} {124}},\ \bibinfo
  {pages} {207001} (\bibinfo {year} {2020})}\BibitemShut {NoStop}%
\bibitem [{\citenamefont {Singh}\ \emph {et~al.}(2017)\citenamefont {Singh},
  \citenamefont {Barker}, \citenamefont {Thamizhavel}, \citenamefont {Paul},
  \citenamefont {Hillier},\ and\ \citenamefont {Singh}}]{singh2017time}%
  \BibitemOpen
  \bibfield  {author} {\bibinfo {author} {\bibfnamefont {D.}~\bibnamefont
  {Singh}}, \bibinfo {author} {\bibfnamefont {J.}~\bibnamefont {Barker}},
  \bibinfo {author} {\bibfnamefont {A.}~\bibnamefont {Thamizhavel}}, \bibinfo
  {author} {\bibfnamefont {D.~M.}\ \bibnamefont {Paul}}, \bibinfo {author}
  {\bibfnamefont {A.}~\bibnamefont {Hillier}},\ and\ \bibinfo {author}
  {\bibfnamefont {R.}~\bibnamefont {Singh}},\ }\bibfield  {title} {\bibinfo
  {title} {Time-reversal symmetry breaking in the noncentrosymmetric
  superconductor {Re}$_6${Hf}: Further evidence for unconventional behavior in
  the $\alpha$-{Mn} family of materials},\ }\href@noop {} {\bibfield  {journal}
  {\bibinfo  {journal} {Physical Review B}\ }\textbf {\bibinfo {volume} {96}},\
  \bibinfo {pages} {180501} (\bibinfo {year} {2017})}\BibitemShut {NoStop}%
\bibitem [{\citenamefont {Pang}\ \emph {et~al.}(2018)\citenamefont {Pang},
  \citenamefont {Nie}, \citenamefont {Wang}, \citenamefont {Singh},
  \citenamefont {Xie}, \citenamefont {Jiang}, \citenamefont {Chen},
  \citenamefont {Singh}, \citenamefont {Smidman},\ and\ \citenamefont
  {Yuan}}]{pang2018fully}%
  \BibitemOpen
  \bibfield  {author} {\bibinfo {author} {\bibfnamefont {G.}~\bibnamefont
  {Pang}}, \bibinfo {author} {\bibfnamefont {Z.}~\bibnamefont {Nie}}, \bibinfo
  {author} {\bibfnamefont {A.}~\bibnamefont {Wang}}, \bibinfo {author}
  {\bibfnamefont {D.}~\bibnamefont {Singh}}, \bibinfo {author} {\bibfnamefont
  {W.}~\bibnamefont {Xie}}, \bibinfo {author} {\bibfnamefont {W.}~\bibnamefont
  {Jiang}}, \bibinfo {author} {\bibfnamefont {Y.}~\bibnamefont {Chen}},
  \bibinfo {author} {\bibfnamefont {R.}~\bibnamefont {Singh}}, \bibinfo
  {author} {\bibfnamefont {M.}~\bibnamefont {Smidman}},\ and\ \bibinfo {author}
  {\bibfnamefont {H.}~\bibnamefont {Yuan}},\ }\bibfield  {title} {\bibinfo
  {title} {Fully gapped superconductivity in single crystals of
  noncentrosymmetric {Re}$_6${Zr} with broken time-reversal symmetry},\
  }\href@noop {} {\bibfield  {journal} {\bibinfo  {journal} {Physical Review
  B}\ }\textbf {\bibinfo {volume} {97}},\ \bibinfo {pages} {224506} (\bibinfo
  {year} {2018})}\BibitemShut {NoStop}%
\bibitem [{\citenamefont {Singh}\ \emph {et~al.}(2018)\citenamefont {Singh},
  \citenamefont {KP}, \citenamefont {Barker}, \citenamefont {Paul},
  \citenamefont {Hillier},\ and\ \citenamefont {Singh}}]{singh2018time}%
  \BibitemOpen
  \bibfield  {author} {\bibinfo {author} {\bibfnamefont {D.}~\bibnamefont
  {Singh}}, \bibinfo {author} {\bibfnamefont {S.}~\bibnamefont {KP}}, \bibinfo
  {author} {\bibfnamefont {J.}~\bibnamefont {Barker}}, \bibinfo {author}
  {\bibfnamefont {D.~M.}\ \bibnamefont {Paul}}, \bibinfo {author}
  {\bibfnamefont {A.}~\bibnamefont {Hillier}},\ and\ \bibinfo {author}
  {\bibfnamefont {R.}~\bibnamefont {Singh}},\ }\bibfield  {title} {\bibinfo
  {title} {Time-reversal symmetry breaking in the noncentrosymmetric
  superconductor {Re}$_6${Ti}},\ }\href@noop {} {\bibfield  {journal} {\bibinfo
   {journal} {Physical Review B}\ }\textbf {\bibinfo {volume} {97}},\ \bibinfo
  {pages} {100505} (\bibinfo {year} {2018})}\BibitemShut {NoStop}%
\bibitem [{\citenamefont {Wilson}\ \emph {et~al.}(1975)\citenamefont {Wilson},
  \citenamefont {Di~Salvo},\ and\ \citenamefont {Mahajan}}]{wilson1975charge}%
  \BibitemOpen
  \bibfield  {author} {\bibinfo {author} {\bibfnamefont {J.~A.}\ \bibnamefont
  {Wilson}}, \bibinfo {author} {\bibfnamefont {F.}~\bibnamefont {Di~Salvo}},\
  and\ \bibinfo {author} {\bibfnamefont {S.}~\bibnamefont {Mahajan}},\
  }\bibfield  {title} {\bibinfo {title} {Charge-density waves and superlattices
  in the metallic layered transition metal dichalcogenides},\ }\href@noop {}
  {\bibfield  {journal} {\bibinfo  {journal} {Advances in Physics}\ }\textbf
  {\bibinfo {volume} {24}},\ \bibinfo {pages} {117} (\bibinfo {year}
  {1975})}\BibitemShut {NoStop}%
\bibitem [{\citenamefont {Wilson}\ \emph {et~al.}(1974)\citenamefont {Wilson},
  \citenamefont {Di~Salvo},\ and\ \citenamefont {Mahajan}}]{wilson1974charge}%
  \BibitemOpen
  \bibfield  {author} {\bibinfo {author} {\bibfnamefont {J.}~\bibnamefont
  {Wilson}}, \bibinfo {author} {\bibfnamefont {F.}~\bibnamefont {Di~Salvo}},\
  and\ \bibinfo {author} {\bibfnamefont {S.}~\bibnamefont {Mahajan}},\
  }\bibfield  {title} {\bibinfo {title} {Charge-density waves in metallic,
  layered, transition-metal dichalcogenides},\ }\href@noop {} {\bibfield
  {journal} {\bibinfo  {journal} {Physical review letters}\ }\textbf {\bibinfo
  {volume} {32}},\ \bibinfo {pages} {882} (\bibinfo {year} {1974})}\BibitemShut
  {NoStop}%
\bibitem [{\citenamefont {Manzeli}\ \emph {et~al.}(2017)\citenamefont
  {Manzeli}, \citenamefont {Ovchinnikov}, \citenamefont {Pasquier},
  \citenamefont {Yazyev},\ and\ \citenamefont {Kis}}]{manzeli20172d}%
  \BibitemOpen
  \bibfield  {author} {\bibinfo {author} {\bibfnamefont {S.}~\bibnamefont
  {Manzeli}}, \bibinfo {author} {\bibfnamefont {D.}~\bibnamefont
  {Ovchinnikov}}, \bibinfo {author} {\bibfnamefont {D.}~\bibnamefont
  {Pasquier}}, \bibinfo {author} {\bibfnamefont {O.~V.}\ \bibnamefont
  {Yazyev}},\ and\ \bibinfo {author} {\bibfnamefont {A.}~\bibnamefont {Kis}},\
  }\bibfield  {title} {\bibinfo {title} {2{D} transition metal
  dichalcogenides},\ }\href@noop {} {\bibfield  {journal} {\bibinfo  {journal}
  {Nature Reviews Materials}\ }\textbf {\bibinfo {volume} {2}},\ \bibinfo
  {pages} {1} (\bibinfo {year} {2017})}\BibitemShut {NoStop}%
\bibitem [{\citenamefont {Liu}\ \emph {et~al.}(2014)\citenamefont {Liu},
  \citenamefont {Li}, \citenamefont {Lu}, \citenamefont {Ang}, \citenamefont
  {Liu},\ and\ \citenamefont {Sun}}]{liu2014coexistence}%
  \BibitemOpen
  \bibfield  {author} {\bibinfo {author} {\bibfnamefont {Y.}~\bibnamefont
  {Liu}}, \bibinfo {author} {\bibfnamefont {L.}~\bibnamefont {Li}}, \bibinfo
  {author} {\bibfnamefont {W.}~\bibnamefont {Lu}}, \bibinfo {author}
  {\bibfnamefont {R.}~\bibnamefont {Ang}}, \bibinfo {author} {\bibfnamefont
  {X.}~\bibnamefont {Liu}},\ and\ \bibinfo {author} {\bibfnamefont
  {Y.}~\bibnamefont {Sun}},\ }\bibfield  {title} {\bibinfo {title} {Coexistence
  of superconductivity and commensurate charge density wave in
  4{Hb}-{Ta}{S}$_{2-x}${Se}$_x$ single crystals},\ }\href@noop {} {\bibfield
  {journal} {\bibinfo  {journal} {Journal of Applied Physics}\ }\textbf
  {\bibinfo {volume} {115}} (\bibinfo {year} {2014})}\BibitemShut {NoStop}%
\bibitem [{\citenamefont {Di~Salvo}\ \emph {et~al.}(1973)\citenamefont
  {Di~Salvo}, \citenamefont {Bagley}, \citenamefont {Voorhoeve},\ and\
  \citenamefont {Waszczak}}]{di1973preparation}%
  \BibitemOpen
  \bibfield  {author} {\bibinfo {author} {\bibfnamefont {F.}~\bibnamefont
  {Di~Salvo}}, \bibinfo {author} {\bibfnamefont {B.}~\bibnamefont {Bagley}},
  \bibinfo {author} {\bibfnamefont {J.}~\bibnamefont {Voorhoeve}},\ and\
  \bibinfo {author} {\bibfnamefont {J.}~\bibnamefont {Waszczak}},\ }\bibfield
  {title} {\bibinfo {title} {Preparation and properties of a new polytype of
  tantalum disulfide (4{Hb}-{Ta}{S}$_2$)},\ }\href@noop {} {\bibfield
  {journal} {\bibinfo  {journal} {Journal of Physics and Chemistry of Solids}\
  }\textbf {\bibinfo {volume} {34}},\ \bibinfo {pages} {1357} (\bibinfo {year}
  {1973})}\BibitemShut {NoStop}%
\bibitem [{\citenamefont {Fei}\ \emph {et~al.}(2022)\citenamefont {Fei},
  \citenamefont {Wu}, \citenamefont {Zhang},\ and\ \citenamefont
  {Yin}}]{fei2022understanding}%
  \BibitemOpen
  \bibfield  {author} {\bibinfo {author} {\bibfnamefont {Y.}~\bibnamefont
  {Fei}}, \bibinfo {author} {\bibfnamefont {Z.}~\bibnamefont {Wu}}, \bibinfo
  {author} {\bibfnamefont {W.}~\bibnamefont {Zhang}},\ and\ \bibinfo {author}
  {\bibfnamefont {Y.}~\bibnamefont {Yin}},\ }\bibfield  {title} {\bibinfo
  {title} {Understanding the mott insulating state in 1${T}$-{Ta}{S}$_2$ and
  1${T}$-{Ta}{Se}$_2$},\ }\href@noop {} {\bibfield  {journal} {\bibinfo
  {journal} {AAPPS Bulletin}\ }\textbf {\bibinfo {volume} {32}},\ \bibinfo
  {pages} {20} (\bibinfo {year} {2022})}\BibitemShut {NoStop}%
\bibitem [{\citenamefont {Law}\ and\ \citenamefont {Lee}(2017)}]{law20171t}%
  \BibitemOpen
  \bibfield  {author} {\bibinfo {author} {\bibfnamefont {K.~T.}\ \bibnamefont
  {Law}}\ and\ \bibinfo {author} {\bibfnamefont {P.~A.}\ \bibnamefont {Lee}},\
  }\bibfield  {title} {\bibinfo {title} {1${T}$-{Ta}{S}$_2$ as a quantum spin
  liquid},\ }\href@noop {} {\bibfield  {journal} {\bibinfo  {journal}
  {Proceedings of the National Academy of Sciences}\ }\textbf {\bibinfo
  {volume} {114}},\ \bibinfo {pages} {6996} (\bibinfo {year}
  {2017})}\BibitemShut {NoStop}%
\bibitem [{\citenamefont {Lin}\ \emph {et~al.}(2020)\citenamefont {Lin},
  \citenamefont {Huang}, \citenamefont {Zhao}, \citenamefont {Qiao},
  \citenamefont {Liu}, \citenamefont {Wu}, \citenamefont {Chen},\ and\
  \citenamefont {Ji}}]{lin2020scanning}%
  \BibitemOpen
  \bibfield  {author} {\bibinfo {author} {\bibfnamefont {H.}~\bibnamefont
  {Lin}}, \bibinfo {author} {\bibfnamefont {W.}~\bibnamefont {Huang}}, \bibinfo
  {author} {\bibfnamefont {K.}~\bibnamefont {Zhao}}, \bibinfo {author}
  {\bibfnamefont {S.}~\bibnamefont {Qiao}}, \bibinfo {author} {\bibfnamefont
  {Z.}~\bibnamefont {Liu}}, \bibinfo {author} {\bibfnamefont {J.}~\bibnamefont
  {Wu}}, \bibinfo {author} {\bibfnamefont {X.}~\bibnamefont {Chen}},\ and\
  \bibinfo {author} {\bibfnamefont {S.-H.}\ \bibnamefont {Ji}},\ }\bibfield
  {title} {\bibinfo {title} {Scanning tunneling spectroscopic study of
  monolayer 1${T}$-{Ta}{S}$_2$ and 1${T}$-{Ta}{Se}$_2$},\ }\href@noop {}
  {\bibfield  {journal} {\bibinfo  {journal} {Nano Research}\ }\textbf
  {\bibinfo {volume} {13}},\ \bibinfo {pages} {133} (\bibinfo {year}
  {2020})}\BibitemShut {NoStop}%
\bibitem [{\citenamefont {Sipos}\ \emph {et~al.}(2008)\citenamefont {Sipos},
  \citenamefont {Kusmartseva}, \citenamefont {Akrap}, \citenamefont {Berger},
  \citenamefont {Forr{\'o}},\ and\ \citenamefont
  {Tuti{\v{s}}}}]{sipos2008mott}%
  \BibitemOpen
  \bibfield  {author} {\bibinfo {author} {\bibfnamefont {B.}~\bibnamefont
  {Sipos}}, \bibinfo {author} {\bibfnamefont {A.~F.}\ \bibnamefont
  {Kusmartseva}}, \bibinfo {author} {\bibfnamefont {A.}~\bibnamefont {Akrap}},
  \bibinfo {author} {\bibfnamefont {H.}~\bibnamefont {Berger}}, \bibinfo
  {author} {\bibfnamefont {L.}~\bibnamefont {Forr{\'o}}},\ and\ \bibinfo
  {author} {\bibfnamefont {E.}~\bibnamefont {Tuti{\v{s}}}},\ }\bibfield
  {title} {\bibinfo {title} {From mott state to superconductivity in
  1${T}$-{Ta}{S}$_2$},\ }\href@noop {} {\bibfield  {journal} {\bibinfo
  {journal} {Nature Materials}\ }\textbf {\bibinfo {volume} {7}},\ \bibinfo
  {pages} {960} (\bibinfo {year} {2008})}\BibitemShut {NoStop}%
\bibitem [{\citenamefont {Ribak}\ \emph {et~al.}(2020)\citenamefont {Ribak},
  \citenamefont {Skiff}, \citenamefont {Mograbi}, \citenamefont {Rout},
  \citenamefont {Fischer}, \citenamefont {Ruhman}, \citenamefont {Chashka},
  \citenamefont {Dagan},\ and\ \citenamefont {Kanigel}}]{ribak2020chiral}%
  \BibitemOpen
  \bibfield  {author} {\bibinfo {author} {\bibfnamefont {A.}~\bibnamefont
  {Ribak}}, \bibinfo {author} {\bibfnamefont {R.~M.}\ \bibnamefont {Skiff}},
  \bibinfo {author} {\bibfnamefont {M.}~\bibnamefont {Mograbi}}, \bibinfo
  {author} {\bibfnamefont {P.}~\bibnamefont {Rout}}, \bibinfo {author}
  {\bibfnamefont {M.}~\bibnamefont {Fischer}}, \bibinfo {author} {\bibfnamefont
  {J.}~\bibnamefont {Ruhman}}, \bibinfo {author} {\bibfnamefont
  {K.}~\bibnamefont {Chashka}}, \bibinfo {author} {\bibfnamefont
  {Y.}~\bibnamefont {Dagan}},\ and\ \bibinfo {author} {\bibfnamefont
  {A.}~\bibnamefont {Kanigel}},\ }\bibfield  {title} {\bibinfo {title} {Chiral
  superconductivity in the alternate stacking compound 4{Hb}-{Ta}{S}$_2$},\
  }\href@noop {} {\bibfield  {journal} {\bibinfo  {journal} {Science advances}\
  }\textbf {\bibinfo {volume} {6}},\ \bibinfo {pages} {eaax9480} (\bibinfo
  {year} {2020})}\BibitemShut {NoStop}%
\bibitem [{\citenamefont {Nayak}\ \emph {et~al.}(2021)\citenamefont {Nayak},
  \citenamefont {Steinbok}, \citenamefont {Roet}, \citenamefont {Koo},
  \citenamefont {Margalit}, \citenamefont {Feldman}, \citenamefont {Almoalem},
  \citenamefont {Kanigel}, \citenamefont {Fiete}, \citenamefont {Yan} \emph
  {et~al.}}]{nayak2021evidence}%
  \BibitemOpen
  \bibfield  {author} {\bibinfo {author} {\bibfnamefont {A.~K.}\ \bibnamefont
  {Nayak}}, \bibinfo {author} {\bibfnamefont {A.}~\bibnamefont {Steinbok}},
  \bibinfo {author} {\bibfnamefont {Y.}~\bibnamefont {Roet}}, \bibinfo {author}
  {\bibfnamefont {J.}~\bibnamefont {Koo}}, \bibinfo {author} {\bibfnamefont
  {G.}~\bibnamefont {Margalit}}, \bibinfo {author} {\bibfnamefont
  {I.}~\bibnamefont {Feldman}}, \bibinfo {author} {\bibfnamefont
  {A.}~\bibnamefont {Almoalem}}, \bibinfo {author} {\bibfnamefont
  {A.}~\bibnamefont {Kanigel}}, \bibinfo {author} {\bibfnamefont {G.~A.}\
  \bibnamefont {Fiete}}, \bibinfo {author} {\bibfnamefont {B.}~\bibnamefont
  {Yan}}, \emph {et~al.},\ }\bibfield  {title} {\bibinfo {title} {Evidence of
  topological boundary modes with topological nodal-point superconductivity},\
  }\href@noop {} {\bibfield  {journal} {\bibinfo  {journal} {Nature physics}\
  }\textbf {\bibinfo {volume} {17}},\ \bibinfo {pages} {1413} (\bibinfo {year}
  {2021})}\BibitemShut {NoStop}%
\bibitem [{\citenamefont {Almoalem}\ \emph
  {et~al.}(2024{\natexlab{a}})\citenamefont {Almoalem}, \citenamefont
  {Feldman}, \citenamefont {Mangel}, \citenamefont {Shlafman}, \citenamefont
  {Yaish}, \citenamefont {Fischer}, \citenamefont {Moshe}, \citenamefont
  {Ruhman},\ and\ \citenamefont {Kanigel}}]{almoalem2024observation}%
  \BibitemOpen
  \bibfield  {author} {\bibinfo {author} {\bibfnamefont {A.}~\bibnamefont
  {Almoalem}}, \bibinfo {author} {\bibfnamefont {I.}~\bibnamefont {Feldman}},
  \bibinfo {author} {\bibfnamefont {I.}~\bibnamefont {Mangel}}, \bibinfo
  {author} {\bibfnamefont {M.}~\bibnamefont {Shlafman}}, \bibinfo {author}
  {\bibfnamefont {Y.~E.}\ \bibnamefont {Yaish}}, \bibinfo {author}
  {\bibfnamefont {M.~H.}\ \bibnamefont {Fischer}}, \bibinfo {author}
  {\bibfnamefont {M.}~\bibnamefont {Moshe}}, \bibinfo {author} {\bibfnamefont
  {J.}~\bibnamefont {Ruhman}},\ and\ \bibinfo {author} {\bibfnamefont
  {A.}~\bibnamefont {Kanigel}},\ }\bibfield  {title} {\bibinfo {title} {The
  observation of $\pi$-shifts in the little-parks effect in
  4{Hb}-{Ta}{S}$_2$},\ }\href@noop {} {\bibfield  {journal} {\bibinfo
  {journal} {Nature Communications}\ }\textbf {\bibinfo {volume} {15}},\
  \bibinfo {pages} {4623} (\bibinfo {year} {2024}{\natexlab{a}})}\BibitemShut
  {NoStop}%
\bibitem [{\citenamefont {Fischer}\ and\ \citenamefont
  {Goryo}(2015)}]{fischer2015symmetry}%
  \BibitemOpen
  \bibfield  {author} {\bibinfo {author} {\bibfnamefont {M.~H.}\ \bibnamefont
  {Fischer}}\ and\ \bibinfo {author} {\bibfnamefont {J.}~\bibnamefont
  {Goryo}},\ }\bibfield  {title} {\bibinfo {title} {Symmetry and gap
  classification of non-symmorphic {Sr}{Pt}{As}},\ }\href@noop {} {\bibfield
  {journal} {\bibinfo  {journal} {Journal of the Physical Society of Japan}\
  }\textbf {\bibinfo {volume} {84}},\ \bibinfo {pages} {054705} (\bibinfo
  {year} {2015})}\BibitemShut {NoStop}%
\bibitem [{\citenamefont {Wang}\ \emph {et~al.}(2024)\citenamefont {Wang},
  \citenamefont {Jiao}, \citenamefont {Meng}, \citenamefont {Zhang},
  \citenamefont {Dai}, \citenamefont {Tu}, \citenamefont {Zhao}, \citenamefont
  {Xin}, \citenamefont {Huang}, \citenamefont {Lei} \emph
  {et~al.}}]{wang2024evidence}%
  \BibitemOpen
  \bibfield  {author} {\bibinfo {author} {\bibfnamefont {H.}~\bibnamefont
  {Wang}}, \bibinfo {author} {\bibfnamefont {Y.}~\bibnamefont {Jiao}}, \bibinfo
  {author} {\bibfnamefont {F.}~\bibnamefont {Meng}}, \bibinfo {author}
  {\bibfnamefont {X.}~\bibnamefont {Zhang}}, \bibinfo {author} {\bibfnamefont
  {D.}~\bibnamefont {Dai}}, \bibinfo {author} {\bibfnamefont {C.}~\bibnamefont
  {Tu}}, \bibinfo {author} {\bibfnamefont {C.}~\bibnamefont {Zhao}}, \bibinfo
  {author} {\bibfnamefont {L.}~\bibnamefont {Xin}}, \bibinfo {author}
  {\bibfnamefont {S.}~\bibnamefont {Huang}}, \bibinfo {author} {\bibfnamefont
  {H.}~\bibnamefont {Lei}}, \emph {et~al.},\ }\bibfield  {title} {\bibinfo
  {title} {Evidence for multiband gapless superconductivity in the topological
  superconductor candidate 4{Hb}-{Ta}{S}$_2$},\ }\href@noop {} {\bibfield
  {journal} {\bibinfo  {journal} {arXiv preprint arXiv:2412.08450}\ } (\bibinfo
  {year} {2024})}\BibitemShut {NoStop}%
\bibitem [{\citenamefont {Meng}\ \emph {et~al.}(2024)\citenamefont {Meng},
  \citenamefont {Fu}, \citenamefont {Pan}, \citenamefont {Tian}, \citenamefont
  {Yan}, \citenamefont {Li}, \citenamefont {Wang}, \citenamefont {Zhang},\ and\
  \citenamefont {Lei}}]{meng2024extreme}%
  \BibitemOpen
  \bibfield  {author} {\bibinfo {author} {\bibfnamefont {F.}~\bibnamefont
  {Meng}}, \bibinfo {author} {\bibfnamefont {Y.}~\bibnamefont {Fu}}, \bibinfo
  {author} {\bibfnamefont {S.}~\bibnamefont {Pan}}, \bibinfo {author}
  {\bibfnamefont {S.}~\bibnamefont {Tian}}, \bibinfo {author} {\bibfnamefont
  {S.}~\bibnamefont {Yan}}, \bibinfo {author} {\bibfnamefont {Z.}~\bibnamefont
  {Li}}, \bibinfo {author} {\bibfnamefont {S.}~\bibnamefont {Wang}}, \bibinfo
  {author} {\bibfnamefont {J.}~\bibnamefont {Zhang}},\ and\ \bibinfo {author}
  {\bibfnamefont {H.}~\bibnamefont {Lei}},\ }\bibfield  {title} {\bibinfo
  {title} {Extreme orbital ab-plane upper critical fields far beyond the pauli
  limit in 4{Hb}-{Ta}({S},{Se})$_2$ bulk crystals},\ }\href@noop {} {\bibfield
  {journal} {\bibinfo  {journal} {Physical Review B}\ }\textbf {\bibinfo
  {volume} {109}},\ \bibinfo {pages} {134510} (\bibinfo {year}
  {2024})}\BibitemShut {NoStop}%
\bibitem [{\citenamefont {Persky}\ \emph {et~al.}(2022)\citenamefont {Persky},
  \citenamefont {Bj{\o}rlig}, \citenamefont {Feldman}, \citenamefont
  {Almoalem}, \citenamefont {Altman}, \citenamefont {Berg}, \citenamefont
  {Kimchi}, \citenamefont {Ruhman}, \citenamefont {Kanigel},\ and\
  \citenamefont {Kalisky}}]{persky2022magnetic}%
  \BibitemOpen
  \bibfield  {author} {\bibinfo {author} {\bibfnamefont {E.}~\bibnamefont
  {Persky}}, \bibinfo {author} {\bibfnamefont {A.~V.}\ \bibnamefont
  {Bj{\o}rlig}}, \bibinfo {author} {\bibfnamefont {I.}~\bibnamefont {Feldman}},
  \bibinfo {author} {\bibfnamefont {A.}~\bibnamefont {Almoalem}}, \bibinfo
  {author} {\bibfnamefont {E.}~\bibnamefont {Altman}}, \bibinfo {author}
  {\bibfnamefont {E.}~\bibnamefont {Berg}}, \bibinfo {author} {\bibfnamefont
  {I.}~\bibnamefont {Kimchi}}, \bibinfo {author} {\bibfnamefont
  {J.}~\bibnamefont {Ruhman}}, \bibinfo {author} {\bibfnamefont
  {A.}~\bibnamefont {Kanigel}},\ and\ \bibinfo {author} {\bibfnamefont
  {B.}~\bibnamefont {Kalisky}},\ }\bibfield  {title} {\bibinfo {title}
  {Magnetic memory and spontaneous vortices in a van der waals
  superconductor},\ }\href@noop {} {\bibfield  {journal} {\bibinfo  {journal}
  {Nature}\ }\textbf {\bibinfo {volume} {607}},\ \bibinfo {pages} {692}
  (\bibinfo {year} {2022})}\BibitemShut {NoStop}%
\bibitem [{\citenamefont {Almoalem}\ \emph
  {et~al.}(2024{\natexlab{b}})\citenamefont {Almoalem}, \citenamefont {Gofman},
  \citenamefont {Nitzav}, \citenamefont {Mangel}, \citenamefont {Feldman},
  \citenamefont {Koo}, \citenamefont {Mazzola}, \citenamefont {Fujii},
  \citenamefont {Vobornik}, \citenamefont {S{\'{}}~anchez Barriga} \emph
  {et~al.}}]{almoalem2024charge}%
  \BibitemOpen
  \bibfield  {author} {\bibinfo {author} {\bibfnamefont {A.}~\bibnamefont
  {Almoalem}}, \bibinfo {author} {\bibfnamefont {R.}~\bibnamefont {Gofman}},
  \bibinfo {author} {\bibfnamefont {Y.}~\bibnamefont {Nitzav}}, \bibinfo
  {author} {\bibfnamefont {I.}~\bibnamefont {Mangel}}, \bibinfo {author}
  {\bibfnamefont {I.}~\bibnamefont {Feldman}}, \bibinfo {author} {\bibfnamefont
  {J.}~\bibnamefont {Koo}}, \bibinfo {author} {\bibfnamefont {F.}~\bibnamefont
  {Mazzola}}, \bibinfo {author} {\bibfnamefont {J.}~\bibnamefont {Fujii}},
  \bibinfo {author} {\bibfnamefont {I.}~\bibnamefont {Vobornik}}, \bibinfo
  {author} {\bibfnamefont {J.}~\bibnamefont {S{\'{}}~anchez Barriga}}, \emph
  {et~al.},\ }\bibfield  {title} {\bibinfo {title} {Charge transfer and
  spin-valley locking in 4{Hb}-{Ta}{S}$_2$},\ }\href@noop {} {\bibfield
  {journal} {\bibinfo  {journal} {npj Quantum Materials}\ }\textbf {\bibinfo
  {volume} {9}},\ \bibinfo {pages} {36} (\bibinfo {year}
  {2024}{\natexlab{b}})}\BibitemShut {NoStop}%
\bibitem [{\citenamefont {Van~Degrift}(1975)}]{van1975tunnel}%
  \BibitemOpen
  \bibfield  {author} {\bibinfo {author} {\bibfnamefont {C.~T.}\ \bibnamefont
  {Van~Degrift}},\ }\bibfield  {title} {\bibinfo {title} {Tunnel diode
  oscillator for 0.001 ppm measurements at low temperatures},\ }\href@noop {}
  {\bibfield  {journal} {\bibinfo  {journal} {Review of Scientific
  Instruments}\ }\textbf {\bibinfo {volume} {46}},\ \bibinfo {pages} {599}
  (\bibinfo {year} {1975})}\BibitemShut {NoStop}%
\bibitem [{\citenamefont {Prozorov}\ and\ \citenamefont
  {Giannetta}(2006)}]{prozorov2006magnetic}%
  \BibitemOpen
  \bibfield  {author} {\bibinfo {author} {\bibfnamefont {R.}~\bibnamefont
  {Prozorov}}\ and\ \bibinfo {author} {\bibfnamefont {R.~W.}\ \bibnamefont
  {Giannetta}},\ }\bibfield  {title} {\bibinfo {title} {Magnetic penetration
  depth in unconventional superconductors},\ }\href@noop {} {\bibfield
  {journal} {\bibinfo  {journal} {Superconductor Science and Technology}\
  }\textbf {\bibinfo {volume} {19}},\ \bibinfo {pages} {R41} (\bibinfo {year}
  {2006})}\BibitemShut {NoStop}%
\bibitem [{\citenamefont {Prozorov}\ \emph {et~al.}(2000)\citenamefont
  {Prozorov}, \citenamefont {Giannetta}, \citenamefont {Carrington},\ and\
  \citenamefont {Araujo-Moreira}}]{prozorov2000meissner}%
  \BibitemOpen
  \bibfield  {author} {\bibinfo {author} {\bibfnamefont {R.}~\bibnamefont
  {Prozorov}}, \bibinfo {author} {\bibfnamefont {R.}~\bibnamefont {Giannetta}},
  \bibinfo {author} {\bibfnamefont {A.}~\bibnamefont {Carrington}},\ and\
  \bibinfo {author} {\bibfnamefont {F.}~\bibnamefont {Araujo-Moreira}},\
  }\bibfield  {title} {\bibinfo {title} {Meissner-london state in
  superconductors of rectangular cross section in a perpendicular magnetic
  field},\ }\href@noop {} {\bibfield  {journal} {\bibinfo  {journal} {Physical
  Review B}\ }\textbf {\bibinfo {volume} {62}},\ \bibinfo {pages} {115}
  (\bibinfo {year} {2000})}\BibitemShut {NoStop}%
\bibitem [{\citenamefont {Prozorov}\ and\ \citenamefont
  {Kogan}(2018)}]{prozorov2018effective}%
  \BibitemOpen
  \bibfield  {author} {\bibinfo {author} {\bibfnamefont {R.}~\bibnamefont
  {Prozorov}}\ and\ \bibinfo {author} {\bibfnamefont {V.~G.}\ \bibnamefont
  {Kogan}},\ }\bibfield  {title} {\bibinfo {title} {Effective demagnetizing
  factors of diamagnetic samples of various shapes},\ }\href@noop {} {\bibfield
   {journal} {\bibinfo  {journal} {Physical review applied}\ }\textbf {\bibinfo
  {volume} {10}},\ \bibinfo {pages} {014030} (\bibinfo {year}
  {2018})}\BibitemShut {NoStop}%
\bibitem [{\citenamefont {Gross}\ \emph {et~al.}(1986)\citenamefont {Gross},
  \citenamefont {Chandrasekhar}, \citenamefont {Einzel}, \citenamefont
  {Andres}, \citenamefont {Hirschfeld}, \citenamefont {Ott}, \citenamefont
  {Beuers}, \citenamefont {Fisk},\ and\ \citenamefont
  {Smith}}]{gross1986anomalous}%
  \BibitemOpen
  \bibfield  {author} {\bibinfo {author} {\bibfnamefont {F.}~\bibnamefont
  {Gross}}, \bibinfo {author} {\bibfnamefont {B.}~\bibnamefont
  {Chandrasekhar}}, \bibinfo {author} {\bibfnamefont {D.}~\bibnamefont
  {Einzel}}, \bibinfo {author} {\bibfnamefont {K.}~\bibnamefont {Andres}},
  \bibinfo {author} {\bibfnamefont {P.}~\bibnamefont {Hirschfeld}}, \bibinfo
  {author} {\bibfnamefont {H.}~\bibnamefont {Ott}}, \bibinfo {author}
  {\bibfnamefont {J.}~\bibnamefont {Beuers}}, \bibinfo {author} {\bibfnamefont
  {Z.}~\bibnamefont {Fisk}},\ and\ \bibinfo {author} {\bibfnamefont
  {J.}~\bibnamefont {Smith}},\ }\bibfield  {title} {\bibinfo {title} {Anomalous
  temperature dependence of the magnetic field penetration depth in
  superconducting {U}{Be}$_{13}$},\ }\href@noop {} {\bibfield  {journal}
  {\bibinfo  {journal} {Zeitschrift f{\"u}r Physik B Condensed Matter}\
  }\textbf {\bibinfo {volume} {64}},\ \bibinfo {pages} {175} (\bibinfo {year}
  {1986})}\BibitemShut {NoStop}%
\bibitem [{\citenamefont {Orlando}\ \emph {et~al.}(1979)\citenamefont
  {Orlando}, \citenamefont {McNiff~Jr}, \citenamefont {Foner},\ and\
  \citenamefont {Beasley}}]{orlando1979critical}%
  \BibitemOpen
  \bibfield  {author} {\bibinfo {author} {\bibfnamefont {T.}~\bibnamefont
  {Orlando}}, \bibinfo {author} {\bibfnamefont {E.}~\bibnamefont {McNiff~Jr}},
  \bibinfo {author} {\bibfnamefont {S.}~\bibnamefont {Foner}},\ and\ \bibinfo
  {author} {\bibfnamefont {M.}~\bibnamefont {Beasley}},\ }\bibfield  {title}
  {\bibinfo {title} {Critical fields, Pauli paramagnetic limiting, and material
  parameters of {Nb}$_3${Sn} and {V}$_3${Si}},\ }\href@noop {} {\bibfield
  {journal} {\bibinfo  {journal} {Physical Review B}\ }\textbf {\bibinfo
  {volume} {19}},\ \bibinfo {pages} {4545} (\bibinfo {year}
  {1979})}\BibitemShut {NoStop}%
\bibitem [{\citenamefont {Fletcher}\ \emph {et~al.}(2007)\citenamefont
  {Fletcher}, \citenamefont {Carrington}, \citenamefont {Diener}, \citenamefont
  {Rodiere}, \citenamefont {Brison}, \citenamefont {Prozorov}, \citenamefont
  {Olheiser},\ and\ \citenamefont {Giannetta}}]{fletcher2007penetration}%
  \BibitemOpen
  \bibfield  {author} {\bibinfo {author} {\bibfnamefont {J.}~\bibnamefont
  {Fletcher}}, \bibinfo {author} {\bibfnamefont {A.}~\bibnamefont
  {Carrington}}, \bibinfo {author} {\bibfnamefont {P.}~\bibnamefont {Diener}},
  \bibinfo {author} {\bibfnamefont {P.}~\bibnamefont {Rodiere}}, \bibinfo
  {author} {\bibfnamefont {J.-P.}\ \bibnamefont {Brison}}, \bibinfo {author}
  {\bibfnamefont {R.}~\bibnamefont {Prozorov}}, \bibinfo {author}
  {\bibfnamefont {T.}~\bibnamefont {Olheiser}},\ and\ \bibinfo {author}
  {\bibfnamefont {R.}~\bibnamefont {Giannetta}},\ }\bibfield  {title} {\bibinfo
  {title} {Penetration depth study of superconducting gap structure of
  2{H}-{Nb}{Se}$_2$},\ }\href@noop {} {\bibfield  {journal} {\bibinfo
  {journal} {Physical review letters}\ }\textbf {\bibinfo {volume} {98}},\
  \bibinfo {pages} {057003} (\bibinfo {year} {2007})}\BibitemShut {NoStop}%
\bibitem [{\citenamefont {Carrington}\ and\ \citenamefont
  {Manzano}(2003)}]{carrington2003magnetic}%
  \BibitemOpen
  \bibfield  {author} {\bibinfo {author} {\bibfnamefont {A.}~\bibnamefont
  {Carrington}}\ and\ \bibinfo {author} {\bibfnamefont {F.}~\bibnamefont
  {Manzano}},\ }\bibfield  {title} {\bibinfo {title} {Magnetic penetration
  depth of {Mg}{B}$_2$},\ }\href@noop {} {\bibfield  {journal} {\bibinfo
  {journal} {Physica C: Superconductivity}\ }\textbf {\bibinfo {volume}
  {385}},\ \bibinfo {pages} {205} (\bibinfo {year} {2003})}\BibitemShut
  {NoStop}%
\bibitem [{\citenamefont {Wang}\ \emph {et~al.}(2001)\citenamefont {Wang},
  \citenamefont {Plackowski},\ and\ \citenamefont {Junod}}]{wang2001specific}%
  \BibitemOpen
  \bibfield  {author} {\bibinfo {author} {\bibfnamefont {Y.}~\bibnamefont
  {Wang}}, \bibinfo {author} {\bibfnamefont {T.}~\bibnamefont {Plackowski}},\
  and\ \bibinfo {author} {\bibfnamefont {A.}~\bibnamefont {Junod}},\ }\bibfield
   {title} {\bibinfo {title} {Specific heat in the superconducting and normal
  state (2-300 {K}, 0-16 {T}), and magnetic susceptibility of the 38 {K}
  superconductor {Mg}{B}$_2$: evidence for a multicomponent gap},\ }\href@noop
  {} {\bibfield  {journal} {\bibinfo  {journal} {Physica C: Superconductivity}\
  }\textbf {\bibinfo {volume} {355}},\ \bibinfo {pages} {179} (\bibinfo {year}
  {2001})}\BibitemShut {NoStop}%
\bibitem [{\citenamefont {Yang}\ \emph {et~al.}(2001)\citenamefont {Yang},
  \citenamefont {Lin}, \citenamefont {Li}, \citenamefont {Hsu}, \citenamefont
  {Liu}, \citenamefont {Li}, \citenamefont {Yu},\ and\ \citenamefont
  {Jin}}]{yang2001order}%
  \BibitemOpen
  \bibfield  {author} {\bibinfo {author} {\bibfnamefont {H.}~\bibnamefont
  {Yang}}, \bibinfo {author} {\bibfnamefont {J.-Y.}\ \bibnamefont {Lin}},
  \bibinfo {author} {\bibfnamefont {H.}~\bibnamefont {Li}}, \bibinfo {author}
  {\bibfnamefont {F.}~\bibnamefont {Hsu}}, \bibinfo {author} {\bibfnamefont
  {C.-J.}\ \bibnamefont {Liu}}, \bibinfo {author} {\bibfnamefont {S.-C.}\
  \bibnamefont {Li}}, \bibinfo {author} {\bibfnamefont {R.-C.}\ \bibnamefont
  {Yu}},\ and\ \bibinfo {author} {\bibfnamefont {C.-Q.}\ \bibnamefont {Jin}},\
  }\bibfield  {title} {\bibinfo {title} {Order parameter of {Mg}{B}$_2$: a
  fully gapped superconductor},\ }\href@noop {} {\bibfield  {journal} {\bibinfo
   {journal} {Physical review letters}\ }\textbf {\bibinfo {volume} {87}},\
  \bibinfo {pages} {167003} (\bibinfo {year} {2001})}\BibitemShut {NoStop}%
\bibitem [{\citenamefont {Bouquet}\ \emph {et~al.}(2001)\citenamefont
  {Bouquet}, \citenamefont {Wang}, \citenamefont {Fisher}, \citenamefont
  {Hinks}, \citenamefont {Jorgensen}, \citenamefont {Junod},\ and\
  \citenamefont {Phillips}}]{bouquet2001phenomenological}%
  \BibitemOpen
  \bibfield  {author} {\bibinfo {author} {\bibfnamefont {F.}~\bibnamefont
  {Bouquet}}, \bibinfo {author} {\bibfnamefont {Y.}~\bibnamefont {Wang}},
  \bibinfo {author} {\bibfnamefont {R.}~\bibnamefont {Fisher}}, \bibinfo
  {author} {\bibfnamefont {D.}~\bibnamefont {Hinks}}, \bibinfo {author}
  {\bibfnamefont {J.}~\bibnamefont {Jorgensen}}, \bibinfo {author}
  {\bibfnamefont {A.}~\bibnamefont {Junod}},\ and\ \bibinfo {author}
  {\bibfnamefont {N.}~\bibnamefont {Phillips}},\ }\bibfield  {title} {\bibinfo
  {title} {Phenomenological two-gap model for the specific heat of
  {Mg}{B}$_2$},\ }\href@noop {} {\bibfield  {journal} {\bibinfo  {journal}
  {Europhysics Letters}\ }\textbf {\bibinfo {volume} {56}},\ \bibinfo {pages}
  {856} (\bibinfo {year} {2001})}\BibitemShut {NoStop}%
\bibitem [{\citenamefont {Kumar~Nayak}\ \emph {et~al.}(2023)\citenamefont
  {Kumar~Nayak}, \citenamefont {Steinbok}, \citenamefont {Roet}, \citenamefont
  {Koo}, \citenamefont {Feldman}, \citenamefont {Almoalem}, \citenamefont
  {Kanigel}, \citenamefont {Yan}, \citenamefont {Rosch}, \citenamefont
  {Avraham} \emph {et~al.}}]{kumar2023first}%
  \BibitemOpen
  \bibfield  {author} {\bibinfo {author} {\bibfnamefont {A.}~\bibnamefont
  {Kumar~Nayak}}, \bibinfo {author} {\bibfnamefont {A.}~\bibnamefont
  {Steinbok}}, \bibinfo {author} {\bibfnamefont {Y.}~\bibnamefont {Roet}},
  \bibinfo {author} {\bibfnamefont {J.}~\bibnamefont {Koo}}, \bibinfo {author}
  {\bibfnamefont {I.}~\bibnamefont {Feldman}}, \bibinfo {author} {\bibfnamefont
  {A.}~\bibnamefont {Almoalem}}, \bibinfo {author} {\bibfnamefont
  {A.}~\bibnamefont {Kanigel}}, \bibinfo {author} {\bibfnamefont
  {B.}~\bibnamefont {Yan}}, \bibinfo {author} {\bibfnamefont {A.}~\bibnamefont
  {Rosch}}, \bibinfo {author} {\bibfnamefont {N.}~\bibnamefont {Avraham}},
  \emph {et~al.},\ }\bibfield  {title} {\bibinfo {title} {First-order quantum
  phase transition in the hybrid metal--mott insulator transition metal
  dichalcogenide 4{Hb}-{Ta}{S}$_2$},\ }\href@noop {} {\bibfield  {journal}
  {\bibinfo  {journal} {Proceedings of the National Academy of Sciences}\
  }\textbf {\bibinfo {volume} {120}},\ \bibinfo {pages} {e2304274120} (\bibinfo
  {year} {2023})}\BibitemShut {NoStop}%
\bibitem{zhou_2025_15688310}
Yuwei Zhou.
\newblock "Nodeless superconductivity in 4{Hb}-{Ta}{S}$_2$ with broken time
  reversal symmetry" [{D}ata set], 2025.
\newblock \doi {10.5281/zenodo.15688310}.

\end{thebibliography}
\end{document}